\documentclass[twocolumn,english,prl,amssymb,aps,superscriptaddress,showpacs,twocolumn,amsmath,showkeys,floatfix]{revtex4-1}
\usepackage[T1]{fontenc}
\usepackage[latin9]{inputenc}
\usepackage[active]{srcltx}
\usepackage{textcomp}
\usepackage{amsmath}
\usepackage{graphicx}
\usepackage{color}
\usepackage{esint}
\usepackage{caption}

\makeatletter
\usepackage{babel}

\makeatother

\begin{document}

\title {Optimization of laser dynamics for active stabilization of DF--VECSELs dedicated to cesium CPT clocks}
\author{Grégory Gredat}
\affiliation{Universit\'e Paris-Saclay, CNRS, ENS Paris-Saclay, CentraleSup\'elec, LuMIn, Gif-sur-Yvette, France}
\author{Hui Liu}
\affiliation{Universit\'e Paris-Saclay, CNRS, ENS Paris-Saclay, CentraleSup\'elec, LuMIn, Gif-sur-Yvette, France}
\affiliation{Institute of Photonics and Photon Technology, Northwest University, Xi'An 710069, China}
\author{Jérémie Cotxet}
\affiliation{Thales Research \& Technology, Palaiseau, France}
\author{François Tricot}
\affiliation{Thales Research \& Technology, Palaiseau, France}
\author{Ghaya Baili}
\affiliation{Thales Research \& Technology, Palaiseau, France}
\author{François Gutty}
\affiliation{Thales Research \& Technology, Palaiseau, France}
\author{Fabienne Goldfarb}
\affiliation{Universit\'e Paris-Saclay, CNRS, ENS Paris-Saclay, CentraleSup\'elec, LuMIn, Gif-sur-Yvette, France}
\author{Isabelle Sagnes}
\affiliation{Centre de Nanosciences et Nanotechnologie (C2N), CNRS, Université Paris-Saclay, Palaiseau, France}
\author{Fabien Bretenaker}
\affiliation{Universit\'e Paris-Saclay, CNRS, ENS Paris-Saclay, CentraleSup\'elec, LuMIn, Gif-sur-Yvette, France}
\affiliation{Light and Matter Physics Group, Raman Research Institute, Bangalore 560080, India}

\begin{abstract} 
We report the implementation and performance of a double servo--loop  for intensity and phase--difference active stabilization of a dual--frequency vertical external--cavity surface--emitting laser (DF--VECSEL)  for  coherent population trapping (CPT) of cesium atoms in the framework of compact atomic clocks. In--phase fully correlated pumping of the two laser modes is identified as the best scheme for intensity noise reduction, and an analytical model allows the optimization of the active stabilization strategy. Optical phase--locking the beat--note to a local oscillator leads to a phase noise level below -103~dBc/Hz at 100~Hz from the carrier. The laser contribution to the short--term frequency stability of the clock is predicted to be compatible with a targeted Allan deviation below $\sigma_y = 5\,\times 10^{-13}$ over one second.
\end{abstract}


\maketitle
\section{Introduction}

The SI unit of time is based on the ground state hyperfine transition frequency of cesium 133 atoms around 9.2 GHz. This definition has been enabled by the possibility to build more and more accurate and stable cesium--based frequency standards \cite{Diddams1318}. Cesium fountains are nowadays primary standards with outstanding properties. 
Meanwhile, secondary standards have been developed and optical lattice clocks have reach such high performances \cite{nemitz2016frequency} that they can be used to measure gravitational time dilation for a given elevation on earth \cite{takano2016geopotential} as predicted by the theory of general relativity. However, for embedded technologies such as global navigation satellite systems (GNSS), communication networks, remote and multi-platform sensing, reduction of the volume of the clock is mandatory. Commercial chip--scale atomic clocks made with rubidium or cesium atoms exhibit low cost and small size, but their performances are not sufficient for many of these applications. A trade-off between size and performance is necessary. There is thus a need for clocks reaching quite good short--term stability, i.~e.  a targeted relative Allan deviation of $\sigma_y = 5 \times 10^{-13}$ at one second averaging time, in a volume of the order of few liters. One way to improve the compactness of such clocks is to implement an all--optical interrogation of atoms. With the use of a cesium double lambda transition that provides well contrasted Ramsey fringes \cite{Zanon:2005}, state--of--the--art CPT--based cesium clocks exhibit a good relative frequency stability of $3.2 \times 10^{-13}$ over one second \cite{Danet:2014}. These clocks use a configuration of the local oscillator, which is not compact enough (around 100\,liters) for GNSS outage solutions, multistatic radar systems, radar electronic warfare, etc. Besides, low--noise dual--frequency external cavity semiconductor lasers have been developed \cite{Baili:2009}. Such dual-frequency lasers innately offer the advantage of optimal modulation depth for the production of optically carried RF signals through the beatnote between two balanced orthogonal polarizations \cite{Alouini:2001}. A promising way to optically probe the double lambda transition, compatible with improved compactness, has thus arised with lin$\perp$lin DF--VECSELs at 852~nm along the $\mathrm{D_2}$ line of cesium \cite{Dumont:2014}. However, the laser noise contribution to the Allan standard deviation is dominant and  special attention has to be paid to  noise suppression strategies. In particular, the observed relative intensity noise (RIN) level of -115 dB/Hz has been shown to limit the clock relative stability to $\sigma_y = 1.6 \times 10^{-12}$ over one second. It is consequently mandatory to reduce the RIN and the phase noise of the beat--note of such a laser by implementing active servo--loop controls. For example, DF--laser beat--note stabilization has already been implemented in DBR fiber lasers with optical feedback \cite{Liang:14} or optical phase--locked loops (OPLL) \cite{Guionie:18}. However, noise investigations in free running DF-VECSELs operating at cesium clock wavelength have shown that the intensity noise and the beat-note phase noise of the free running lasers mainly originate from the intensity noise of the pump laser  \cite{Liu:2018}. The way this pump noise is transferred to the intensity and beat-note phase noises of the DF-VECSEL depends mainly on i) the laser dynamical behavior through the nonlinear coupling constant between the two laser modes and ii) the correlations between the noises of the pump power seen by the two modes. In particular, reducing the noises of the DF-VECSEL requires both a strong in--phase correlation of the pump noises seen by the two modes and a low nonlinear coupling between these modes. Such conditions can be met using a fully--correlated multi--mode pumping architecture with a relatively large spatial separation between the two modes in the active medium \cite{Gredat:2018}.

Based on these preceding results on the free running laser noises, the aim of the present paper is to investigate the role of the  DF--VECSEL dynamics and of the pump  noise correlation in the optimization of the active stabilization loops. These loops aim at reducing the  intensity noise and the beat-note phase noise of the DF-VECSEL below the levels necessary to reach the targeted clock performances. 

The article is organized as follows. Section\,\ref{sec:1} presents the formalism necessary to model the nonlinear dynamics of the laser responsible for the propagation of the pump noise to the DF-VECSEL noise. In particular, we introduce the laser coupled Langevin rate equations and the correlation between the pump noises seen by the two modes, allowing to predict the output noise correlations between the two emitted modes at 852~nm. Two pumping architectures allow  to explore various situations for both the correlation of the pump noises and the coupling strength between the modes. These architectures thus lead to different laser dynamics, which are detailed. We then use this formalism in Sec.\,\ref{sec:2} to decide which situation is the most favorable one to achieve an efficient intensity stabilization. To this aim, a servo-loop model is established for intensity feedback with an error signal made with either only one detected mode or with the sum of the two modes. The results of the predictions are compared with the experiment for the different servo-locking strategies. The model  then allows us to improve the loop filter, which leads to the demonstration of a relative intensity noise below -140~dB/Hz at 10~kHz Fourier frequency. The beat--note phase stabilization is implemented in Sec.\,\ref{sec:3}. Thanks to the pumping architecture reported in \cite{Gredat:2018}, the relevant beat-note phase--noise bandwidth lies within the bandwidth of an OPLL. An electro--optic crystal is thus inserted inside the external cavity for feedback and we demonstrate a beat-note phase noise level below -103 dBc/Hz at 100 Hz from the carrier. A simple model of the loop is also built for noise reduction optimization. Section\,\ref{sec:4}  evaluates the resulting contribution of the DF--VECSEL noise to the short--term stability of the CPT--clock through the Dick effect \cite{dick:87}.  

\section{Laser dynamics and control of the noise correlation} \label{sec:1}

\begin{figure}[h]
    \centering
    \includegraphics[width=.49\textwidth]{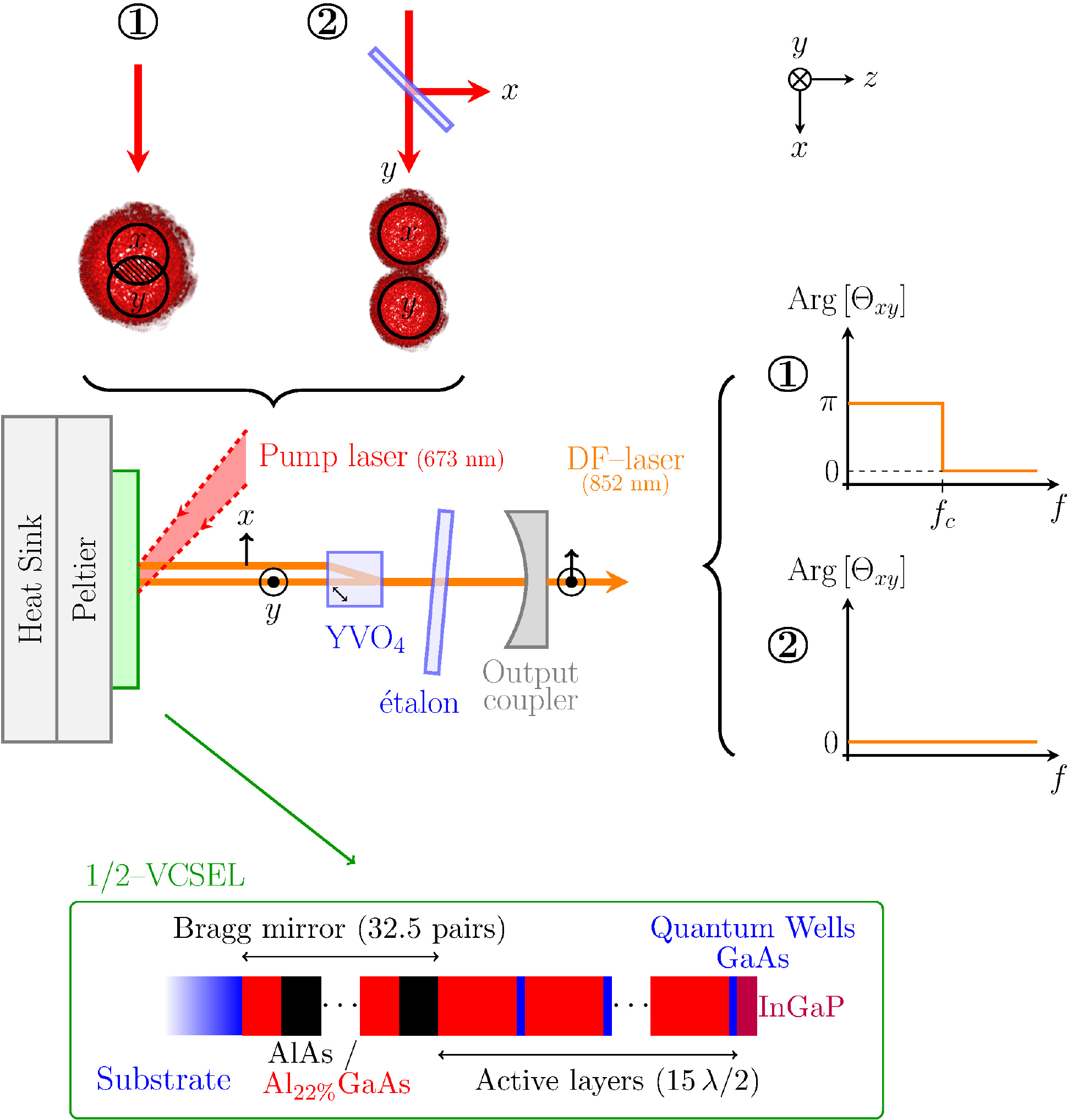}    \caption{Dual-frequency VECSEL oscillating at 852~nm, implemented with two different pumping schemes labeled\textcircled{\raisebox{-1pt}{1}} and \textcircled{\raisebox{-1pt}{2}}. Scheme \textcircled{\raisebox{-1pt}{1}} (resp. \textcircled{\raisebox{-1pt}{2}}) leads to a large (resp. small) value of the coupling constant $C$ between the two modes and a small (resp. large) correlation amplitude $\eta$ between the pump noises seen by the two modes.}
    \label{fig1}
\end{figure}

The VECSEL cavity sketched in Fig.\,\ref{fig1} can emit two orthogonal linearly polarized modes (labeled $x$ and $y$) around 852~nm. This linear cavity is composed of a semiconductor chip, which is based on a distributed Bragg reflector and active quantum wells, and a concave output coupler. The semiconductor chip is referred to as a 1/2--VCSEL and is made of AlGaAs ternary compounds. On top of the Bragg mirror deposited on a GaAs substrate, the 1/2-VCSEL contains seven GaAs quantum wells embedded in $\mathrm{Al_{0.22}Ga_{0.78}As}$ barriers. These quantum wells are providing gain. The 1/2-VCSEL structure is glued to a Peltier cooler, which is itself bonded to a heat sink. We pump the quantum wells with a multimode--fibered pump laser diode operating at  673~nm. An etalon is inserted inside the cavity to strengthen the mode stability. We then insert a 1~mm thick positive uniaxial birefringent crystal of $\mathrm{YVO_4}$, which is cut at $45^\circ$ off its optical axis. It creates a walk--off between the ordinary  and the extraordinary polarized modes. This reduces the competition between these two modes, thus enabling the dual--frequency operation necessary to excite the CPT resonance of cesium.

Such a laser obeys class-A dynamics \cite{Baili:2009}, thanks to the fact that the cavity photon lifetimes $\tau_x$ and $\tau_y$  for the two modes are much longer than the carrier lifetime $\tau$. Even if this leads to a natural reduction of the intensity noises of the two modes, further active intensity noise suppression is necessary for metrological application of these lasers. An active intensity feedback must consequently be implemented through a control on the pump laser driver input, as is going to be investigated in the next section. As a consequence, a deep understanding of the very details of the pump--to--laser noise transfer is needed to efficiently optimize this active servo-locking. The 852~nm DF--VECSEL dynamics has been investigated in \cite{Liu:2018} and can be described with the rate equations formalism for the two emitted modes in the presence of gain cross-saturation \cite{De:2013}. The ratios $\xi_{xy}$ and $\xi_{yx}$ of cross-to--self--saturation coefficients are used to model the competition between the $x$- and $y$--polarized modes. As sketched in Fig.\,\ref{fig1}, the pumping regions providing gain to the  $x$ and $y$ modes can indeed partially overlap, depending on the transverse walk--off induced by the birefringent crystal and the mode sizes imposed by the cavity geometry. We call $r_x$ and $r_y$ the excitation ratios of the two modes (ratios of the unsaturated gain and the losses per cavity round-trip). Then, the steady--state numbers of photons  $F_{0x}$ and $F_{0y}$ of the two modes and the associated steady--state  unsaturated carriers numbers $N_{0x}$ and $N_{0y}$ read:
    \begin{eqnarray}
  F_{0i}& =& \frac{1}{\kappa\,\tau}\frac{(r_i-1)-\xi_{ij}\,(r_j -1)}{1-C}\,,\label{eq:1}\\
  \quad N_{0i}&=& \frac{r_i}{\kappa \tau_i}\,,\label{eq:1N1}
  \end{eqnarray}
with $i,j \in \left\lbrace x,y\right\rbrace$, $i \neq j $, $C=\xi_{xy}\,\xi_{yx}$, and $\kappa$  denoting the coupling coefficient between the photons and the carriers. 
The parameter $C$ is referred to as the coupling constant and stable dual--frequency operation requires $C<1$. A linear stability analysis of the laser rate equations shows that the stability of the dual--frequency operation actually imposes a condition on the ratios of cross-to--self--saturation coefficients $\xi_{ij}<(r_i-1)/(r_j-1)$. This condition coincides with the steady-state number of photons of \eqref{eq:1} being positive. 

The fluctuations of the pump powers seen by the two modes induce fluctuations $\delta F_x(t)$ and $\delta F_y(t)$ of their numbers of photons. The pump power fluctuations are modeled by introducing fluctuations $\delta N_{0x}(t)$ and $\delta N_{0y}(t)$ in the unsaturated carrier numbers  $N_{0x}$ and $N_{0y}$ seen by the two modes. Linearization of the laser rate equations \cite{Liu:2018} then leads to the following relation between the Fourier transforms $\widetilde{\delta F_x}$ and $\widetilde{\delta F_y}$ of these photon number fluctuations and the Fourier transforms of $\delta N_{0x}$ and $\delta N_{0y}$: 
\begin{equation}
\label{eq:2}
\left(\begin{array}{c}
\widetilde{\delta F_x}\left(f\right)\\
\widetilde{\delta F_y}\left(f\right)
\end{array}\right) = \left(\begin{array}{cc}
M_{xx}\left(f\right) & M_{xy}\left(f\right)  \\
M_{yx}\left(f\right)  &  M_{yy}\left(f\right)
\end{array}\right) \left(\begin{array}{c}
\widetilde{\delta N_{0x}}\left(f\right)\\
\widetilde{\delta N_{0y}}\left(f\right)
\end{array}\right)\ ,
\end{equation}
where $f$ is the noise frequency and with
\begin{eqnarray}
\label{eq:3-4}
M_{xx}\left(f\right) & = & \dfrac{1}{\Delta\left(f\right)\,\tau}\,\left[\dfrac{1}{\tau_y}-2\,\mathrm{i}\,\pi\,f\,\dfrac{r_y/\tau-2\,\mathrm{i}\,\pi\,f}{\kappa\,F_{y0}}\right] \,,\\
M_{xy}\left(f\right) & = & - \dfrac{\xi_{xy}}{\tau\,\tau_x\,\Delta\left(f\right)}\label{eq:3-4N1}\,,
\end{eqnarray}
and similar expressions for $M_{yy}$ and $M_{yx}$. The denominator $\Delta$ is defined as: 
\begin{eqnarray}
\label{eq:5}
    \Delta\left(f\right) & = & \left[\dfrac{1}{\tau_x}-2\,\mathrm{i}\,\pi\,f\,\dfrac{r_x/\tau-2\,\mathrm{i}\,\pi\,f}{\kappa\,F_{x0}}\right] \nonumber\\
    & & \times \left[\dfrac{1}{\tau_y}-2\,\mathrm{i}\,\pi\,f\,\dfrac{r_y/\tau-2\,\mathrm{i}\,\pi\,f}{\kappa\,F_{y0}}\right]-\dfrac{C}{\tau_x\,\tau_y}\,.
\end{eqnarray}
In such lasers, the pump power is delivered through  a multi--mode fiber, creating a speckle pattern on the semiconductor structure. Moreover, the two laser modes do not necessarily intercept the same region of this speckle, as sketched in Fig.\,\ref{fig1}. As a consequence, they do not exactly experience the same pump noise. More precisely, the pump noises seen by the two laser modes have the same relative intensity noise spectra $\mathrm{RIN_p}\left(f\right)$, but the correlation spectrum $\left\langle \widetilde{\delta N_{0x}}\left(f\right) \widetilde{\delta N_{0y}}^\ast\left(f\right) \right\rangle$ between these pump noises can take different values depending on the details of the pumping architecture. Preceding measurements \cite{De:2015,Liu:2018} have shown that this correlation spectrum can be modeled as:
\begin{equation}
\label{eq:6}
\left\langle \widetilde{\delta N_{0x}}\left(f\right) \widetilde{\delta N_{0y}}^\ast\left(f\right) \right\rangle = \eta \, \mathrm{RIN_p}\left(f\right) \,N_{0x}\,N_{0y}\,\mathrm{e}^{\mathrm{i}\,\Psi}\,,
\end{equation}
where $0\leq\eta\leq1$ describes the amplitude of the pump correlations and $\Psi$ the phase of these correlations. Measurements \cite{De:2015,Liu:2018} show that the pump relative intensity noise $\mathrm{RIN_p}\left(f\right)$ and the pump correlation amplitude $\eta$ and phase $\Psi$ are constant over a 20~MHz bandwidth. Moreover, the phase $\Psi$ is always equal to zero while many values of the correlation amplitude $\eta$ can be reached, depending on the pump architecture. As sketched in Fig.\,\ref{fig1}, we consider here two opposite typical pumping architectures. The one labeled \textcircled{\raisebox{-1pt}{1}} corresponds to a single pump spot, which is large enough to create population for the two modes. This architecture is usually associated with a relatively small spatial separation between the two modes, which still exhibit a strong overlap and thus a relatively large value of $C$. On the contrary, the one labeled \textcircled{\raisebox{-1pt}{2}} corresponds to amplitude division of the pump beam producing two copies of the pump source, which are imaged on the structure to feed the two completely separated modes. In the first architecture, the correlation amplitude $\eta$ and the coupling strength $C$ are typically decreasing mutually when the distance between the two modes is increased. For example, in \cite{Liu:2018}, the couples of laser parameters  $\left(C=0.15, \eta=0.10\right)$ and $\left(C=0.44, \eta=0.45\right)$ are obtained using two different birefringent crystal thicknesses. Reaching a fully correlated pumping with reduced mode competition is though impossible in this configuration. For this reason, the second pumping architecture has been proposed in \cite{Gredat:2018}, leading this time to $\left(C=0.05, \eta=0.98\right)$. Indeed, the two pump copies can be well spatially separated in this configuration while transferring the same pump noise. This architecture has proven to be the best one for minimizing the pump-to-laser noise transfer but we can wonder whether it is still the more efficient one for intensity stabilization using servo-control. To sum up, we compare in the following the merits of two pumping architectures with respect to the efficiency of intensity noise active reduction: the first architecture, which uses only one pump beam, enables to investigate intermediate values for both $C$ and $\eta$ and to vary them simultaneously whereas the second one, with two pump beams, allows the investigation of very low values of $C$ with a maximum pump correlation amplitude $\eta \simeq 1$.

The correlation of \eqref{eq:6} between the pump fluctuations seen by the two modes leads to a partial correlation of the intensity noises of the two modes of the DF--VECSEL. We thus define the correlation spectrum  $\Theta_{xy}\left(f\right)$ between the intensity fluctuations of the $x-$ and $y-$ polarized modes as: 
\begin{equation}
\label{eq:7}
 \Theta_{xy}\left(f\right) = \dfrac{\left\langle \widetilde{\delta F_x}\left(f\right) \widetilde{\delta F_y}^\ast\left(f\right) \right\rangle}{\sqrt{\left\langle \left|\widetilde{\delta F_x}\left(f\right) \right|^2 \right\rangle\,\left\langle \left|\widetilde{\delta F_y}\left(f\right)\right|^2  \right\rangle}}\,.
\end{equation}
A good knowledge of this correlation spectrum is required to optimize the intensity stabilization feedback loop. It is indeed not obvious whether the error signal for the intensity servo-control should be extracted from the measurement of the total laser intensity or from the measurement of the intensity of only one mode. If we assume for example that the intensity noise of one mode is reduced by the servo--loop, we should then figure out how the other mode will behave. In this respect, the correlation between the intensity noises and especially its sign can be expected to play an important role. The two schematized spectra at the right of Fig.\,\ref{fig1} show that the two pumping architectures promote different situations for the correlation between the intensity noises. When cross--saturation is large ($C\lesssim 1$), the intensity fluctuations of the DF-VECSEL modes are anti--correlated at frequencies lower than a cut-off frequency $f_c$, which can be equal to a few hundreds of kHz to a few MHz. Above $f_c$, the fluctuations of the intensities are in--phase correlated. When cross--saturation is low ($C\simeq 0$) and pump correlation is large ($\eta\simeq 1$), only in--phase correlations of the intensity fluctuations of the modes can be observed at all frequencies. 

For a fixed noise frequency $f$ and all other parameters fixed, we can define the so-called inversion pump correlation amplitude $\eta_\mathrm{inv}(f)$ for which the sign of the $x-y$ correlation $\Theta_{xy}\left(f\right)$ reverses.   
Using Eqs.  (\ref{eq:1}-\ref{eq:7}),  $\eta_\mathrm{inv}$ can be expressed as a rational function of the Fourier frequency $f$, with $r_x$, $r_y$, $\xi_{xy}$, $\xi_{yx}$, $\tau_x$, $\tau_y$ and $\tau$ as parameters. In the simpler case where we assume a complete balance between the two modes, i. e., $r_x=r_y\equiv r$, $\tau_x=\tau_y\equiv\tau_\mathrm{cav}$ and $\xi_{xy}=\xi_{yx}\equiv\xi$, it reads:
\begin{equation}
 \label{eq:9}
 \eta_{\mathrm{inv}}\left(f\right) = \dfrac{ 2\,\xi \left[1-\left(1+\xi\right)(f/f_c)^2 \right]}{\left(1+\xi^2\right) -2\,\left(1+\xi\right)\,(f/f_c)^2 +\left(1+\xi\right)^2\,(f/f_c)^4}\,, 
\end{equation}
with
\begin{equation}
2\,\pi\,f_c=\sqrt{\left(r-1\right)/\left(\tau_\mathrm{cav}\,\tau\right)}\ .    
\end{equation}

In our preceding experiments \cite{De:2015,Liu:2018}, the typical values of the characteristic frequency $f_c$ were of the order of  10~MHz. Besides, for the class-A lasers we are dealing with, the noise frequencies for which the transfer of the pump noise to the laser intensity noise is efficient are those that are much smaller than the cavity cut-off frequency $1/(2\pi\tau_\mathrm{cav})$. In the following, we consider frequencies $f$ smaller than 100~kHz, i. e., much smaller than $f_c$. In this case, one can derive an expression of $\eta_{\mathrm{inv}}$, which remains valid when the two mode gain and losses are not balanced, i. e., $r_x\neq r_y$: 
\begin{equation}
 \label{eq:10}
 \eta_{\mathrm{inv}} = \dfrac{r_x^2+r_y^2}{r_x \, r_y} \cdot \dfrac{\sqrt{C}}{1+C}\,.
\end{equation}

\begin{figure}[h]
   \centering
\includegraphics[width=.49\textwidth]{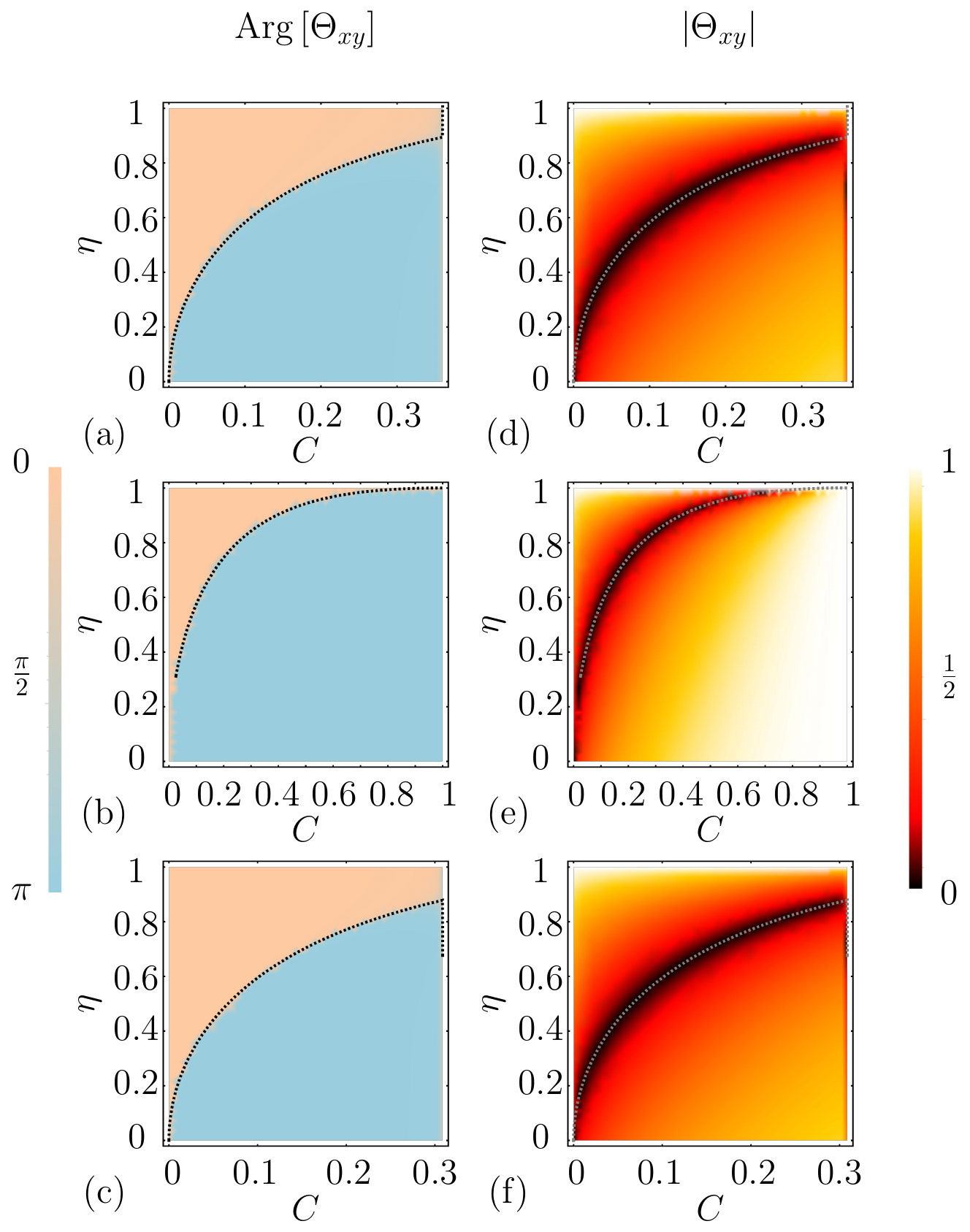}
    \caption{False colour theoretical plots of (a-c) the phase and (d-f) the amplitude of the correlation $\Theta_{xy}$ between the intensity fluctuations of the two modes in the ($C$,$\eta$) plane. The three rows of figures correspond to $r_y/r_x = 0.85$, $r_y/r_x = 1.0$, and $r_y/r_x = 1.3$, respectively. The values of the other parameters are $f=10\,\mathrm{kHz}$,  $\mathrm{RIN_p}=-133\,\mathrm{dB/Hz}$, $\tau = 1\,\mathrm{ns}$, $\tau_x=13\,\mathrm{ns}$, $\tau_y=16\,\mathrm{ns}$ and $r_x=1.6$.
    }
    \label{fig2}
\end{figure}
Figure\,\ref{fig2} represents, using false colours, the evolution of the calculated phase and  modulus of the intensity noise correlation $\Theta_{xy}$ at $f=10\,\mathrm{kHz}$ in the ($C$, $\eta$) plane. The three series of figures correspond to three different values of the ratio $r_x/r_y$ between the excitation rates of the two modes. In Figs.\,\ref{fig2}(a,c,d,f) the range of variation of $C$ is limited to the values for which the two modes can oscillate simultaneously. In Figs.\,\ref{fig2}(b,e), since $r_x=r_y$ the two modes can oscillate simultaneously for all values of $C$ smaller than 1. The dashed line corresponds to \eqref{eq:10} and underlines the change of sign of the correlation. These figures permit to predict the behaviour of the mode intensity noises correlations for any pumping architecture. The role of these correlations on the active stabilization of the intensities is the subject of the following section.

\section{Intensity stabilization}  \label{sec:2}

\begin{figure}[h]
    \centering
    \includegraphics[width=.43\textwidth]{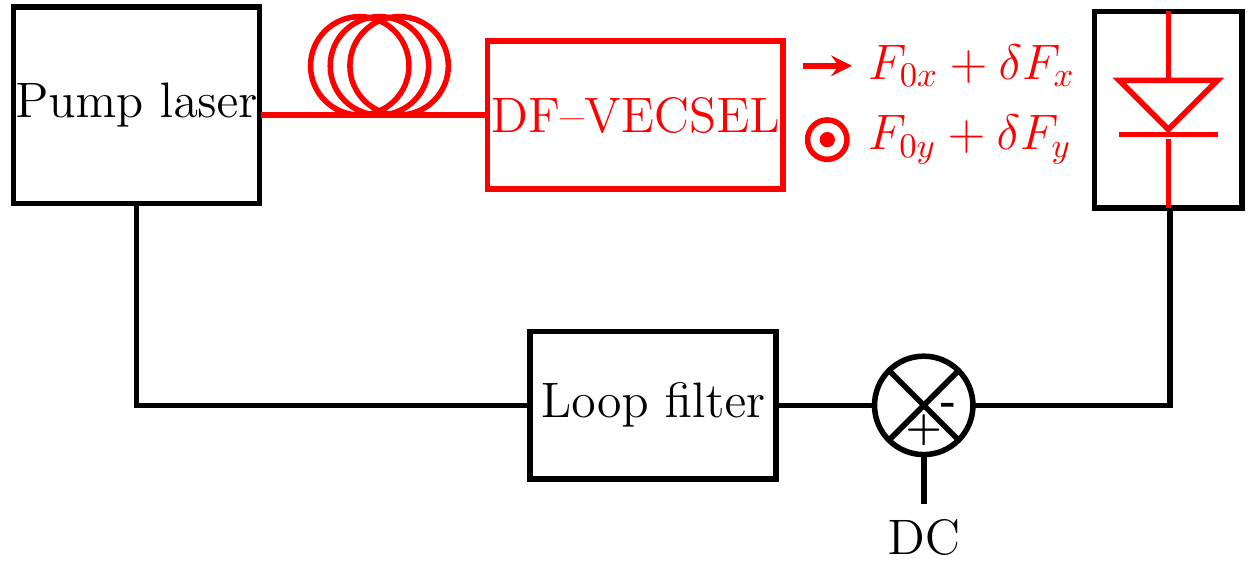}  
    \caption{Servo--loop implementation for RIN reduction. The error signal can be obtained by detecting either a single polarization mode or both modes.}
    \label{fig3}
\end{figure}

In this section, active intensity noise suppression is investigated both theoretically and experimentally. A basic servo--loop scheme for intensity stabilization is shown in Fig.\,\ref{fig3}. After light detection, an error signal is obtained by subtracting the fluctuating intensity with a DC reference. After amplification and filtering, the feedback signal is applied to the modulation input of the pump diode laser driver. This pump current correction then propagates to the DF--VECSEL. Therefore, intensity noise suppression depends on the pump-to-laser transfer and, as a consequence, on the pumping architecture. 

\subsection{Loop model and impact of noise correlations}
\label{subsec:3A}

The error signal used for the intensity servo-loop is filtered with a proportional-integral controller. The corresponding transfer function for unitary gain is denoted as $H_\mathrm{PI}\left(f\right)$ and is characterized by a corner frequency $f_\mathrm{PI}$. The pump laser driver modulation input is used for the pump current feedback. In our case, measurements show that the transfer function from input current correction to pump irradiation can be approximated by a second--order low--pass filter model. This transfer function is denoted as $H_\mathrm{LP}\left(f\right)$ and its associated parameters are its cut--off frequency $f_\mathrm{LP}$ and its quality factor $Q$. The total servo-loop transfer function $H$ is thus
\begin{equation}
    H\left(f\right)~=~G \cdot H_\mathrm{LP}\left(f\right)\cdot H_\mathrm{PI}\left(f\right)\ ,
\end{equation}
where $G$ is a global loop gain factor.

In the following, two cases are distinguished depending on whether the intensity of only one laser or the two laser modes is detected to create the feedback signal. 

\subsubsection{Feedback based on detection of only one mode}

In this section, we assume that the feedback signal is generated from the detection of only one mode, namely the $x$-polarized one. Since the correction applied to the pump does not discriminate between the pumped regions corresponding to the $x-$ and $y-$polarized modes, the propagation of the pump intensity variations to the DF-VECSEL intensity fluctuations described in section \ref{sec:1} comes into the picture. The closed-loop relative intensity noise (RIN) of the $x-$polarized mode, denoted $\mathrm{RIN}^\mathrm{lock}_{x}(f)$, is related to the free--running RIN of the $x-$polarized mode $\mathrm{RIN}_{x}(f)$ through the following equation:    
\begin{equation}
\label{eq:11}
\mathrm{RIN}^\mathrm{lock}_{x}\left(f\right)    =  \left|\dfrac{1}{1+H\left(f\right)\cdot \left( M_{xx}\left(f\right)+M_{xy}\left(f\right) \right)}\right|^2  \mathrm{RIN}_{x}\left(f\right)\,,
\end{equation}
where the matrix elements $M_{xx}$ and $M_{xy}$ are given by Eqs.\,(\ref{eq:3-4}) and (\ref{eq:3-4N1}), respectively.

This servo-loop stabilizing the intensity of the $x-$polarized mode also affects the intensity noise on the $y-$polarized mode, leading to the following closed-loop RIN spectrum:
\begin{multline}
\label{eq:12}
\mathrm{RIN}^\mathrm{lock}_{y}\left(f\right)  =  \mathrm{RIN}_{y}\left(f\right) +\left|G_{xy}\left(f\right)\right|^2   \cdot \left(\dfrac{F_{0x}}{F_{0y}} \right)^2 \cdot \mathrm{RIN}_{x}\left(f\right)\\
 -2\,\mathrm{Re}\left[G_{xy}\left(f\right) \Theta_{xy}\left(f\right)\right] \cdot \left(\dfrac{F_{0x}}{F_{0y}} \right) \cdot \sqrt{\mathrm{RIN}_{x}\left(f\right)\mathrm{RIN}_{y}\left(f\right)}\,,
\end{multline}
with
\begin{equation}  
\label{eq:13}
G_{xy}\left(f\right)=\dfrac{H\left(f\right)\cdot \left( M_{yy}\left(f\right)+M_{yx}\left(f\right) \right)}{1+H\left(f\right)\cdot \left( M_{xx}\left(f\right)+M_{xy}\left(f\right) \right)}.
\end{equation}

\begin{figure}[h]
    \centering
    \includegraphics[width=0.495\columnwidth]{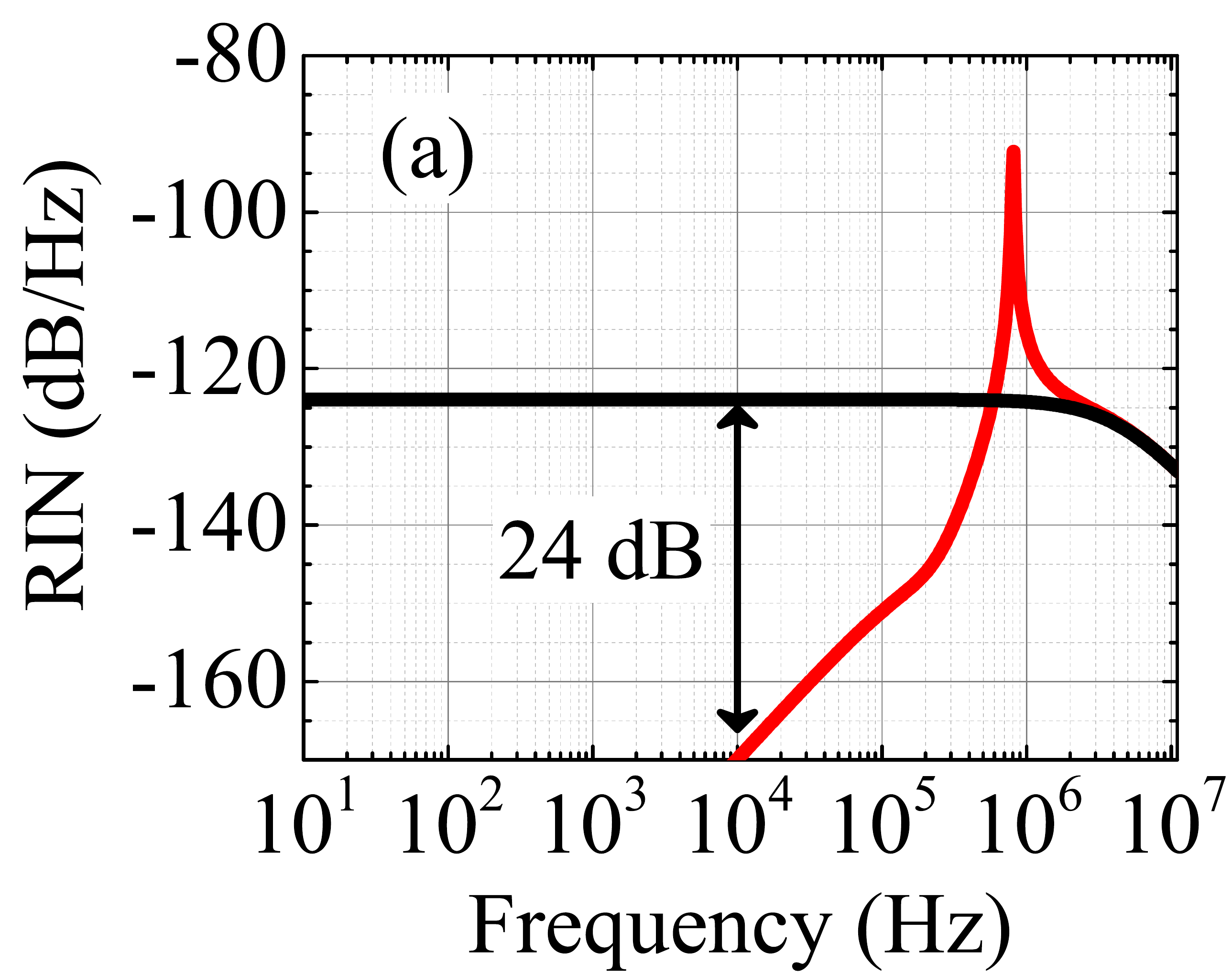}  \includegraphics[width=0.495\columnwidth]{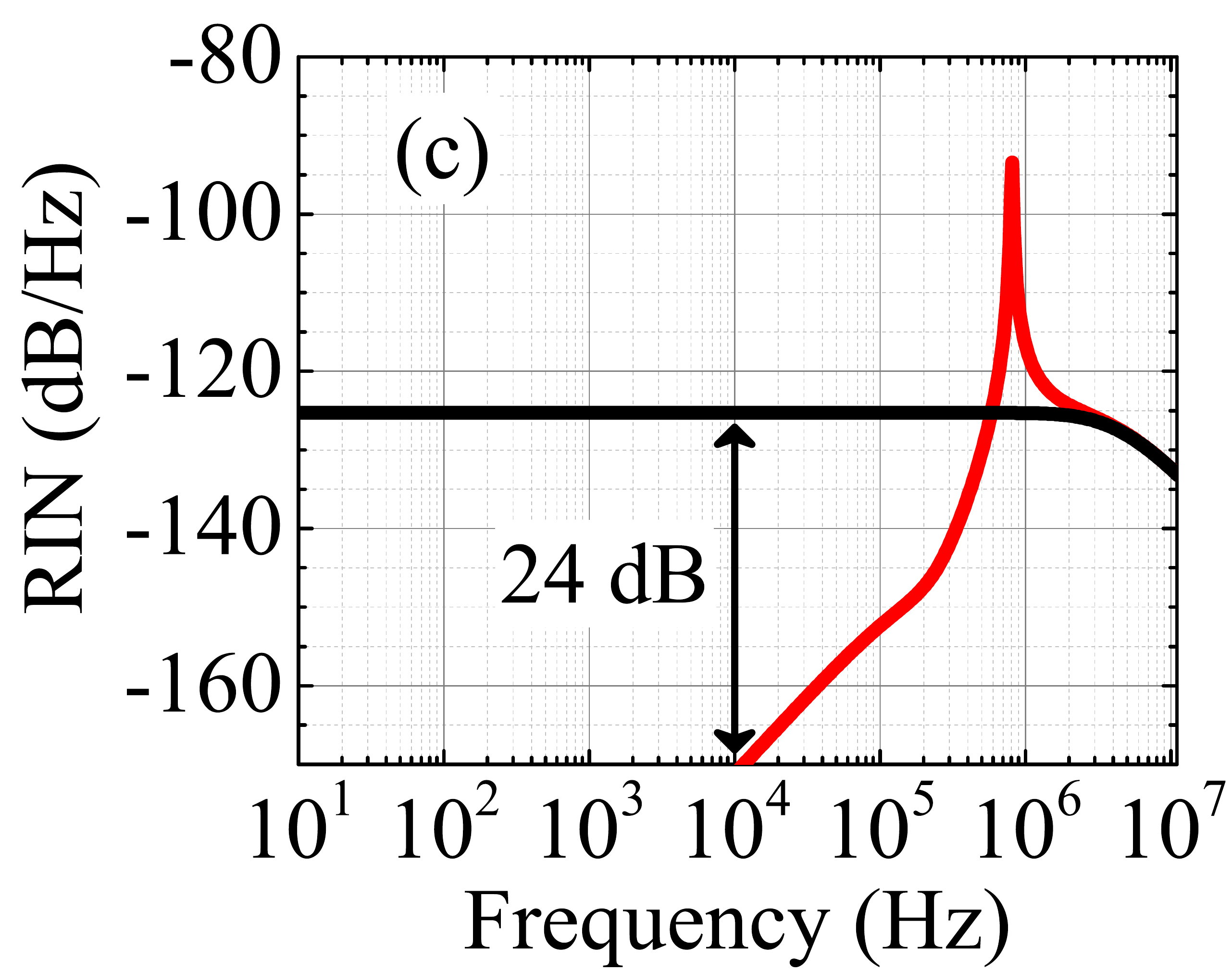} \includegraphics[width=0.495\columnwidth]{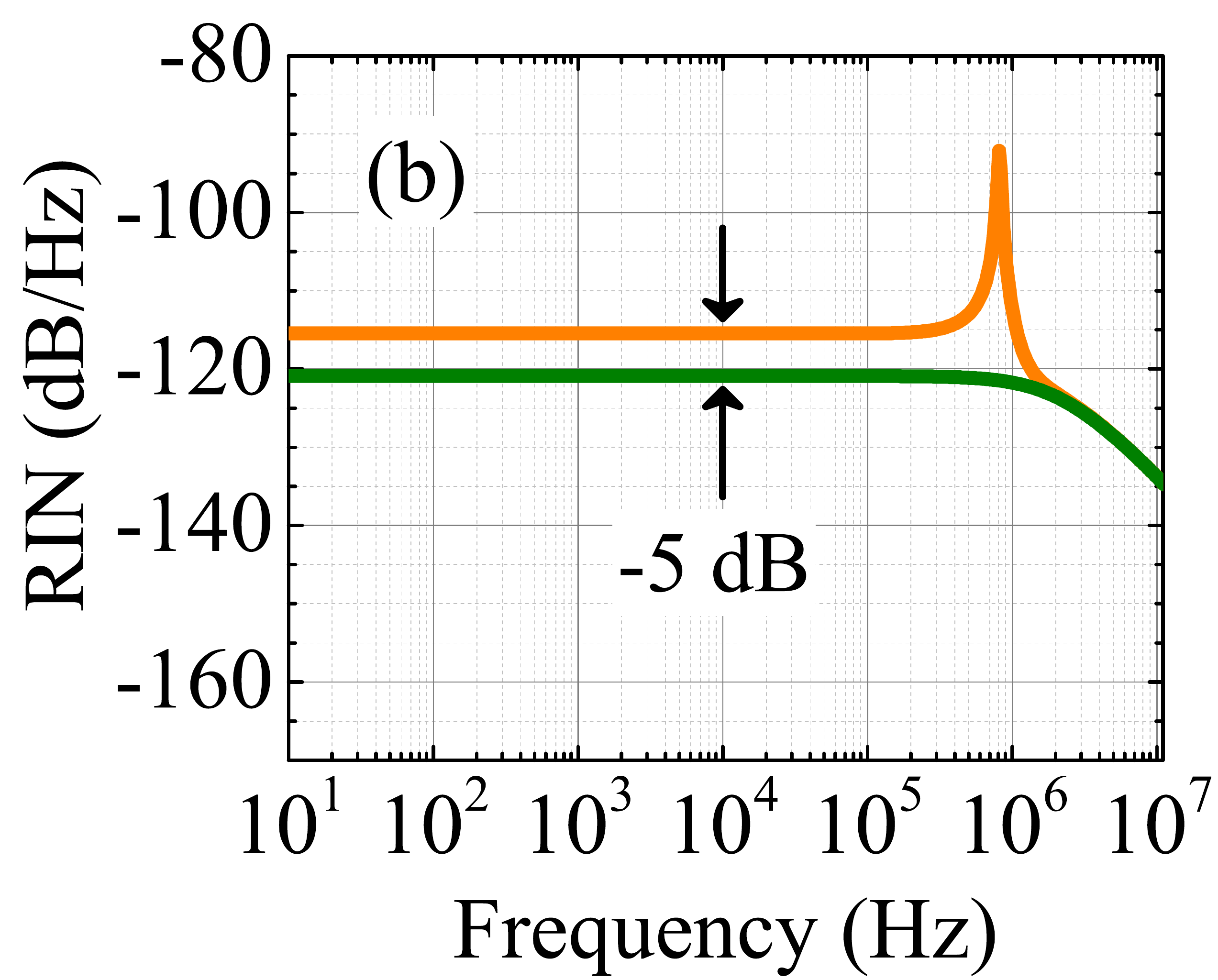} \includegraphics[width=0.495\columnwidth]{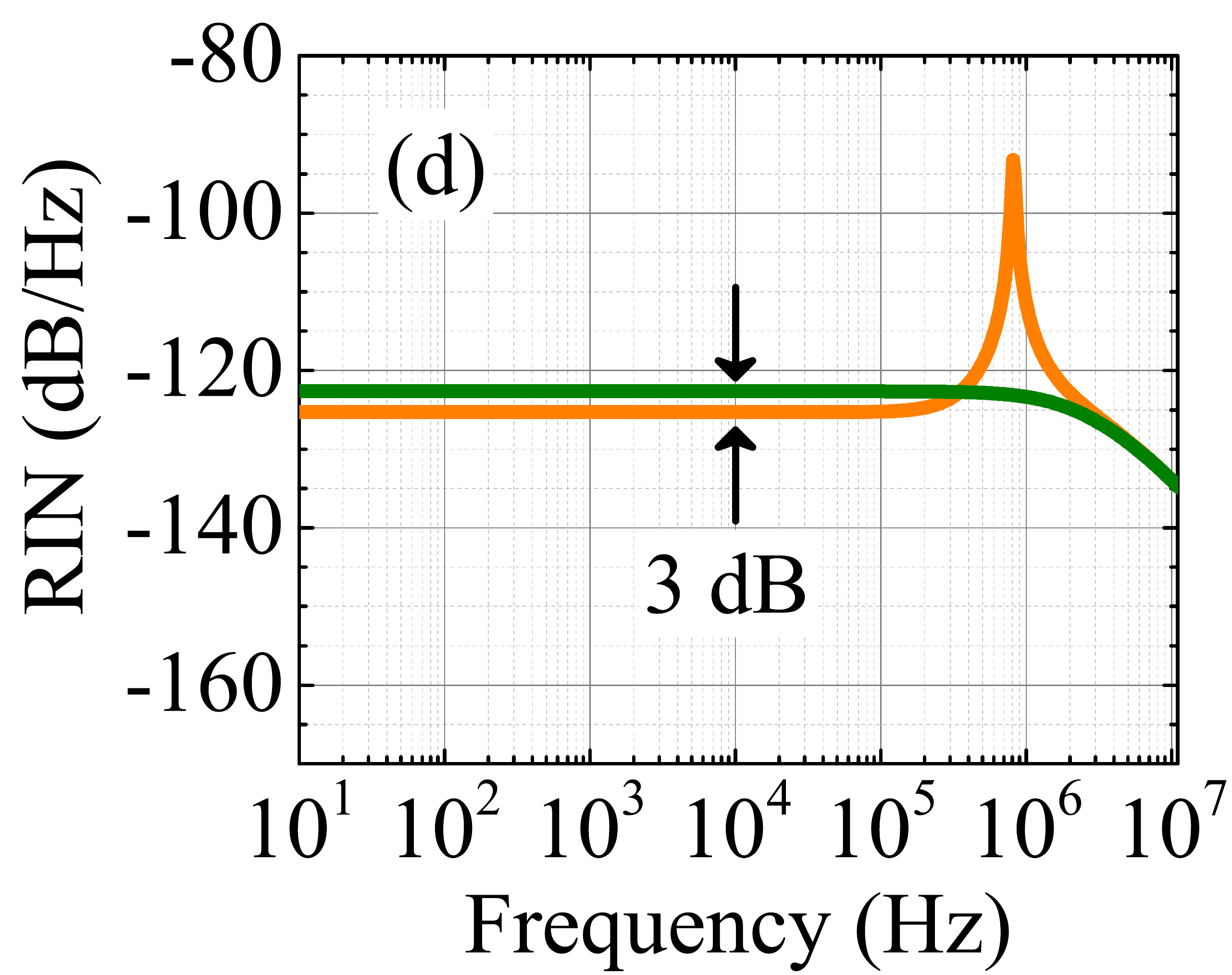}
    \caption{Theoretical RIN spectra for (a,c) the $x-$polarized or (b,d) the $y-$polarized laser mode, either in free-running condition or when the intensity of the $x-$polarized mode is stabilized. The  black arrow stresses the RIN change at 10~kHz between the free running laser (black line for $x$, green line for $y$) and the locked laser (red line for $x$, orange line for $y$). (a,b) $\eta=0.4$; (c,d) $\eta=0.98$. Servo-loop parameters: $G=0.8$, $f_\mathrm{PI}=200\,\mathrm{kHz}$, $f_\mathrm{LP}=250\,\mathrm{kHz}$, $Q=1$.  Laser parameters: $\mathrm{RIN_p}=-133\,\mathrm{dB/Hz}$, $\tau_x = 13\,\mathrm{ns}$, $\tau_y = 16\,\mathrm{ns}$, $\tau=1\,\mathrm{ns}$, $r_x = 1.6$, $r_y = 1.45$, and $C=0.05$.}
    \label{fig4}
\end{figure}

 Typical closed-loop RIN spectra obtained from Eqs.\,(\ref{eq:11}) and (\ref{eq:12}) are shown in Fig.\,\ref{fig4} for different values of the pumping noise correlation amplitude $\eta$. The values of the parameters that we take correspond to the measurements that will be described in section \ref{ExpRIN}. As can be seen in Figs.\,\ref{fig4}(a) and \ref{fig4}(c) for $\eta=0.4$ and $\eta=0.98$, respectively, the RIN of the $x$-polarized mode is heavily reduced on a 300~kHz bandwidth after the loop is closed. This reduction is independent of the value of $\eta$ because only the $x$-polarized mode is used for feedback. In both cases, a RIN decrease of 24~dB is obtained at 10~kHz frequency thanks to the servo-loop. The peak around 800~kHz corresponds to the frequency for which the phase shift of the transfer function $H\left(f\right)\cdot \left( M_{xx}\left(f\right)+M_{xy}\left(f\right) \right)$ reaches $\pi$: the loop then amplifies the noise instead of decreasing it. We will see in section \ref{Leadlag} below how one can mitigate this peak. 
 
 Using Eq.\,(\ref{eq:12}), one can also simulate the modification of the RIN spectrum of the $y-$polarized mode when the RIN of the $x-$ polarized one is reduced by the loop. The results are reproduced for $C=0.05$ in Figs.\,\ref{fig4}(b) and \ref{fig4}(d) for $\eta=0.4$ and $\eta=0.98$, respectively.  One can see that the RIN of the $y$-polarized mode at 10~kHz can be either increased or decreased by the action of the servo-loop, depending on the value of $\eta$. For a given amount of cross-saturation (which is low in the case of Fig.\,\ref{fig4} for which $C~=~0.05$), strong  correlations between the pump fluctuations is more favorable for a mutual reduction of the intensity noises of the two modes. This suggests that the configuration based on two pump beams, labeled as \textcircled{\raisebox{-1pt}{2}} in Fig.\,\ref{fig1}, is the more appropriate one for intensity stabilization. 
 
 The origin of these predictions can be understood by looking at the different terms in the right-hand side of \eqref{eq:12}. The first term is the free-running RIN of the $y-$polarized mode. The second term corresponds to an addition of noise associated with a fraction of the RIN of the $x-$polarized mode. Finally, the fact that the total RIN of the $y-$polarized mode is increased or decreased depends on the sign of the last term in \eqref{eq:12}, and more precisely on the sign of $\mathrm{Re}\left[G_{xy}\left(f\right)\Theta_{xy}\left(f\right)\right]$. A decrease of the RIN of the $y-$polarized mode by stabilizing the intensity of the $x-$polarized one is thus possible only if this quantity is positive. 
 
 
  If all the parameters of $x-$ and $y-$polarized modes are very close (same pumping ratios, same photon lifetimes, same ratios of cross-to-self saturation coefficients and RINs), we have $G_{xy}~\sim~1$ for large loop gains and low noise frequencies. Then, in the case of independent mode noises $\Theta_{xy}\simeq0$, \eqref{eq:12} shows that the active reduction of the intensity noise of the $x$-polarized mode can only deteriorate the noise of the $y$-polarized one by 3~dB. On the contrary, when the modulus of $\Theta_{xy}$ becomes close to 1 (strong correlation between the mode noises) the last term in \eqref{eq:12} plays an important role. Indeed, then,  strong in--phase correlation of the noises of the modes allows an efficient reduction of the $y$-intensity noise, whereas an anti-phase correlation  can lead to an increase up to 6~dB of the RIN of the $y-$ polarized mode when the loop is closed.

 In the low frequency range, within the servo-locking bandwidth, we have $M_{ii}>0$ and $M_{ij}<0$ with $i \neq j \in \left\lbrace x,y \right\rbrace$, while the sign of the correlation between the RINs of the two modes depends on the pumping parameters as detailed in section \ref{sec:1}.  The term $\mathrm{Re}\left[G_{xy}\left(f\right)\Theta_{xy}\left(f\right)\right]$ can be positive in this frequency range either (i) when the mode RINs are in-phase correlated and $\xi_{yx} < \tau_y/\tau_x$, or (ii) when the mode RINs are anti-phase correlated and $\xi_{yx} > \tau_y/\tau_x$, which means that the cross-saturation term $\vert M_{yx}\vert$ is larger than $M_{yy}$. Let us suppose that the mode with the shortest photon lifetime is the $x-$polarized one, such that $\tau_x \leq \tau_y$. This induces a limit value $C_{\mathrm{lim}}$ for the coupling constant, above which cross--saturation becomes the dominant mechanism of intensity noise for the $y$-polarized mode, given by: 
 \begin{equation}
     C_{\mathrm{lim}}=\left(\tau_x/\tau_y\right)^2\ .\label{Clim}
 \end{equation}
 When the intensity noise along the $x$-polarization is suppressed, the modulation induced by the cross--saturation term in the noise of the $y$-polarization is decreased. As a consequence, decreasing $x$-mode intensity noise can lead to a reduction of the noise in $y$-mode for 
 $C>C_{\mathrm{lim}}$ even though the RINs of $x$ and $y$ are anti-phase correlated. 
 
\begin{figure}[h!]
    \centering
    \includegraphics[width=.32\textwidth]{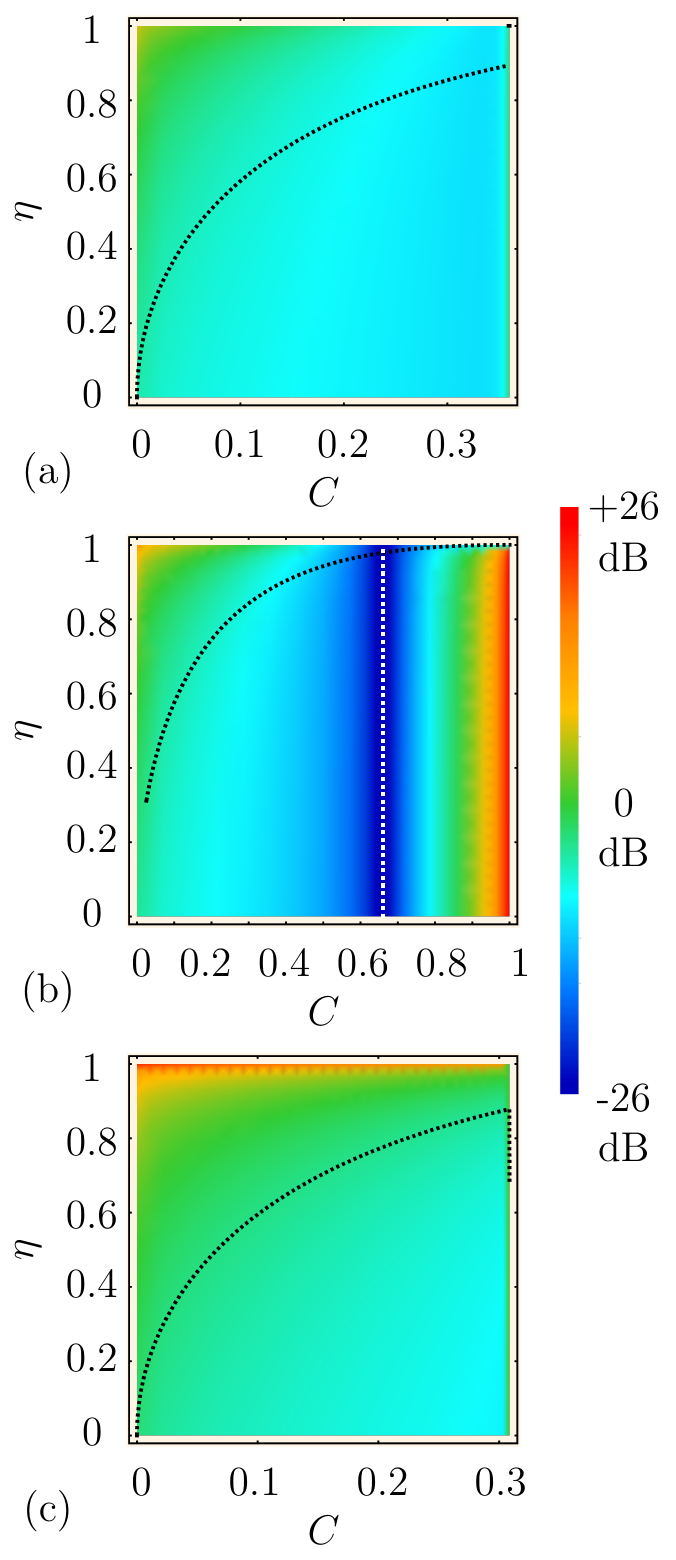}
    \caption{Predicted reduction of the RIN of the $y$ mode at 10~kHz in presence of the loop stabilizing the intensity of the $x$ mode, plotted in the  ($C$,$\eta$) plane. (a) $r_y/r_x = 0.85$, (b) $r_y/r_x = 1.0$, (c) $r_y/r_x  = 1.3$. Loop parameters: $G=0.08$, $f_\mathrm{PI}=200\,\mathrm{kHz}$,  $f_\mathrm{LP}=250\,\mathrm{kHz}$, $Q=1$. Laser parameters: $\mathrm{RIN_p}=-133\,\mathrm{dB/Hz}$, $\tau_x= 13\,\mathrm{ns}$, $\tau_y = 16\,\mathrm{ns}$, $\tau=1\,\mathrm{ns}$ and $r_x= 1.6$. This corresponds to a 24~dB decrease of the RIN of the $x$ mode at 10 kHz. Black dashed line: $\eta_{\mathrm{inv}}$ given by \eqref{eq:10}. White dashed line: $C_\mathrm{lim}=(\tau_x/\tau_y)^2$ (see text).}
    \label{fig5}
\end{figure}
Figure\,\ref{fig5} reproduces in false colors the RIN reduction experienced by the  $y$-polarized mode at 10~kHz frequency when the loop that stabilizes the $x-$polarized mode is closed. This reduction is plotted in the ($C$, $\eta$) plane for different values of $r_x/r_y$. Positive numbers (yellow and red colors) correspond to a reduction of RIN while negative numbers (blue colors) correspond to an increase. In this figure we have taken $\tau_y>\tau_x$. 

Let us first discuss Figs.\,\ref{fig5}(a) and \ref{fig5}(c) for which $r_x\neq r_y$. In both cases the horizontal axis is limited to the values of $C$ for which the two modes oscillate simultaneously: if $C$ is too large, the stronger mode kills the other one. As expected from the preceding discussion, reduction of the RIN of the $y$-polarized mode thanks to the stabilization of the intensity of the $x-$polarized one is possible when $\eta$ is rather large and $C$ rather weak, i. e., when the laser parameters are well above the dashed line that represents the value $\eta_{\mathrm{inv}}$ discussed above. This can be understood by the fact that decreasing the RIN for the $x$ mode in the presence of strong pump correlations and low competition also leads to pump noise reduction in the pump area seen by the $y$ mode. This then leads to the efficient decrease of the noise of the $y$ mode seen at the top left corner of Figs.\,\ref{fig5}(a) and \ref{fig5}(c).

Moreover, comparison of Fig.\,\ref{fig5}(a) with Fig.\,\ref{fig5}(c) shows that a better reduction of the RIN of the $y$ mode can be achieved for $\eta\simeq1$ when $r_y>r_x$ than when $r_y<r_x$. The RIN stabilization loop is thus more efficient when it is based on the detection of the intensity of the weakest mode.

A new regime can be observed in Fig.\,\ref{fig5}(b) for which the excitation ratio of the two modes are equal: $r_x=r_y$. Then, the two modes can oscillate simultaneously for values of $C$ as large as one (notice the horizontal scale). The white vertical line corresponds to the value $C_{\mathrm{lim}}$ given by \eqref{Clim}, which corresponds to the worse degradation of the RIN of the $y$ mode. The situation improves for $C>C_{\mathrm{lim}}$ thanks to the predominant role of the last term in \eqref{eq:12}, as discussed above. Finally, for $C$ very close to 1, one can achieve a very strong reduction of the RIN of the $y$ mode, as evidenced by the red color along the right vertical axis of Fig.\,\ref{fig5}(b). In these conditions, cross-gain saturation becomes so strong that an active reduction of the noise of the $x$ mode also strongly reduces the noise of the $y$ mode. However, such a situation is not very easy to achieve experimentally and is detrimental to the RIN transfer from the pump to the VECSEL \cite{Liu:2018}.

  \subsubsection{Feedback based on the detection of both modes} 
  
  We now assume that the feedback signal is generated from the detection of the total laser power, which is proportional to the total number of photons $F_x+F_y$. This loop will thus reduce the relative intensity noise of the total laser power according to
   \begin{equation}
   \label{eq:14}
\mathrm{RIN}^\mathrm{lock}_{x+y}\left(f\right)    =  \left|\dfrac{1}{1+H\left(f\right)\cdot \sum_{ij} M_{ij}\left(f\right)}\right|^2\mathrm{RIN}_{x+y}\left(f\right)\,,
\end{equation}
where $\mathrm{RIN}_{x+y}$ and $\mathrm{RIN}^\mathrm{lock}_{x+y}$ are the total laser RIN in the open and closed loop cases, respectively. The RIN obtained for one mode after locking the servo--loop, with $i \neq j \in \left\lbrace x,y \right\rbrace$, reads~:
\begin{multline}
\label{eq:15}
\mathrm{RIN}^\mathrm{lock}_{i}\left(f\right)    =  \left|1-K_{ij}\right|^2\mathrm{RIN}_{i}
+ \left|K_{ij} \right|^2 \cdot \left(\dfrac{F_{0j}}{F_{0i}} \right)^2 \cdot \mathrm{RIN}_{j}  \\
-2\, \mathrm{Re}\left[ \left(1-K_{ij}\right) \cdot K_{ij}^\ast \cdot  \Theta_{ij} \right] \cdot \left(\dfrac{F_{0j}}{F_{0i}}\right) \cdot \sqrt{\mathrm{RIN}_{i} \cdot \mathrm{RIN}_{j}}\,,
\end{multline}
where
\begin{equation}
\label{eq:16}
K_{ij}\left(f\right)=\dfrac{H\left(f\right) \cdot \left(M_{ii}\left(f\right)+M_{ij}\left(f\right) \right)}{1+H\left(f\right)\cdot \sum_{ij} M_{ij}\left(f\right)}\,.
\end{equation}
 In the case where the $x$ and $y$ modes have similar parameters (pumping ratios, photon lifetimes, ratios of cross-to-self saturation coefficients, and RINs), one  has $K_{ij} \sim 1/2$ for large loop gains and low noise frequencies. For independent $x$ and $y$ modes ($\Theta_{xy}\simeq 0$), \eqref{eq:15} shows that reducing the intensity noise based on detection of the total intensity reduces the noise of each mode by 3~dB. This equation also shows that in--phase correlation of the mode noises allows an efficient reduction of the intensity noises of both modes whereas in the case anti-phase correlations, closing the loop leaves the noise unchanged.
  
   \begin{figure}[h]
    \centering
    \includegraphics[width=.49\textwidth]{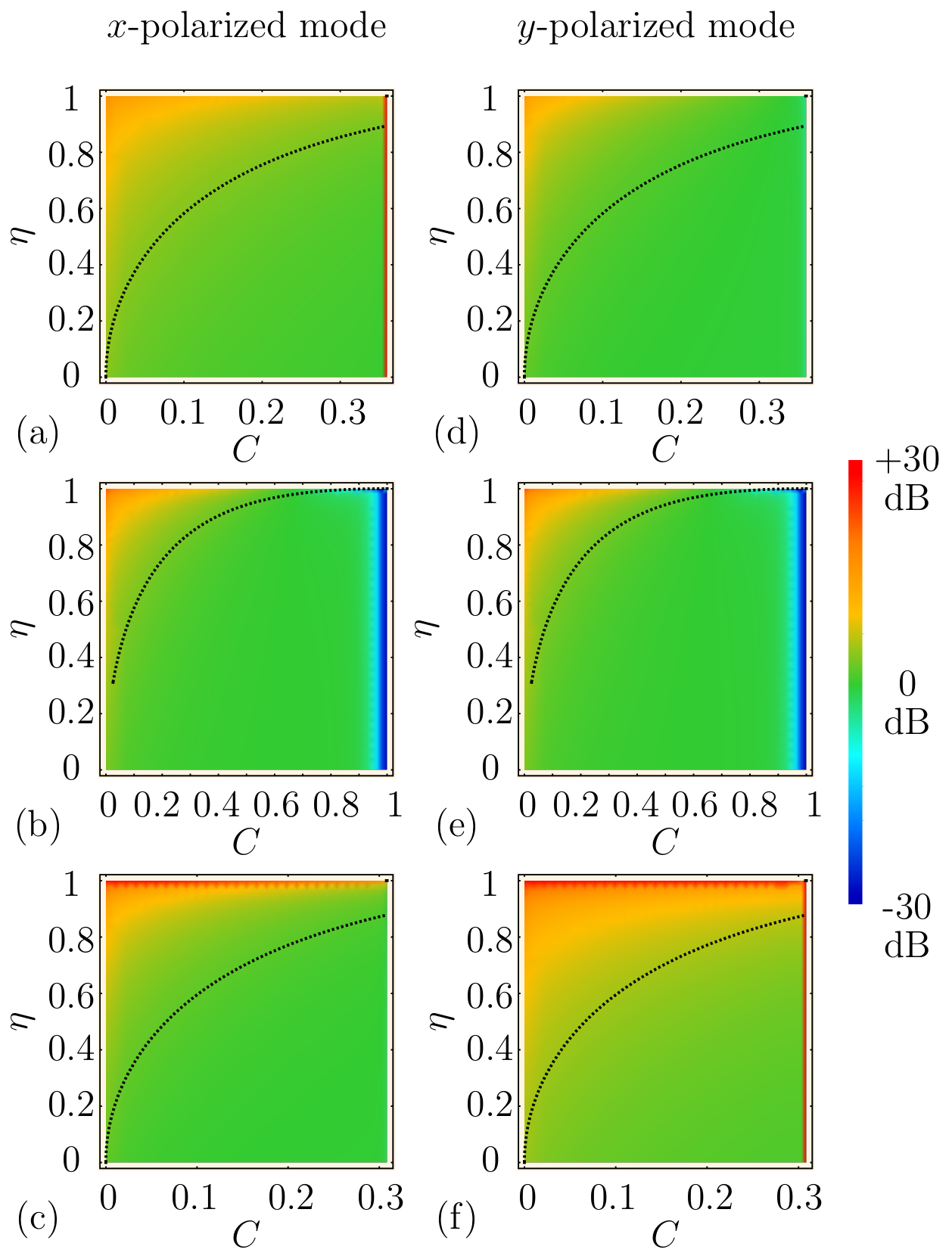}
    \caption{Predicted closed-loop reduction of the RIN at 10~kHz for (a),(b),(c) the $x$- and (d),(e),(f) the $y$-polarized mode.(a),(d): $r_y/r_x = 0.85$; (b),(e): $r_y/r_x = 1.0$; (c),(f):  $r_y/r_x = 1.3$. Loop parameters: $G=0.08$, $f_\mathrm{PI}=200\,\mathrm{kHz}$,  $f_\mathrm{LP}=250\,\mathrm{kHz}$, $Q=1$. Laser parameters: $\mathrm{RIN_p}=-133\,\mathrm{dB/Hz}$, $\tau_x = 13\,\mathrm{ns}$, $\tau_y = 16\,\mathrm{ns}$, $\tau=1\,\mathrm{ns}$ and $r_x = 1.6$. Black dashed line: inversion correlation amplitude $\eta_{\mathrm{inv}}$ given by \eqref{eq:10}.}
    \label{fig6}
\end{figure}
  Figure \ref{fig6} reproduces the closed-loop RIN reduction at 10~kHz for both $x$- and $y$-polarized modes in the ($C$, $\eta$) plane for (b,e) $r_x=r_y$ and (a,c,d,f) $r_x\neq r_y$. In the latter case, the range of variation of $C$ is limited to the region where the two modes oscillate simultaneously. As compared with Fig.\,\ref{fig5}, we observe that the regions in which the RIN is reduced are larger than when  the error signal was extracted from the detection of one mode only. The positive RIN reduction regions (yellow and red colors) of the  ($C$, $\eta$) plane coincide with the in-phase correlations regions of Fig.\,\ref{fig2}, i. e., above the  $\eta_\mathrm{inv}\left(C\right)$ black dashed curve. The stronger mode is the one whose noise is more efficiently reduced by the servo-loop. In the anti-phase correlation region of the plane, the intensity noise of the weaker mode is not affected by the servo--loop while the noise of the stronger one is slightly suppressed for intermediate values of $C$.

  When the coupling constant $C$ is close to 1, which is achievable when $r_x=r_y$ as in Figs.\,\ref{fig6}(b,e), the noises of the $x$- and $y$-polarized modes become fully anti-phase correlated, as shown in Fig.\,\ref{fig2}. Indeed cross-saturation promotes anti-phase noise correlations. The increase of the intensity noises of the two modes evidenced by the blue color close to $C=1$ in Figs.\ref{fig6}(b,e) is due to the fact the servo-control creates an unbalance between the two modes whose effect on the intensities is amplified by the strong competition.
  
    
To conclude this section, our model predicts that, contrary to the case where the feedback is achieved with only one mode, the feedback signal obtained from the sum of the modes enables a mutual reduction of the intensity noises of both modes for a broader range of realistic experimental situations (putting apart the special case of $C \simeq 1$ for which, anyway, the global noise properties are bad). In all cases, the pumping architecture based on two separate spots (labeled as \textcircled{\raisebox{-1pt}{2}} in Fig.\,\ref{fig1}) appears as the best one for efficient noise reduction. This remains to be experimentally verified.  

\subsection{Experimental investigation of intensity stabilization}\label{ExpRIN}

\begin{figure}[h]
\centering
 \includegraphics[width=.495\linewidth]{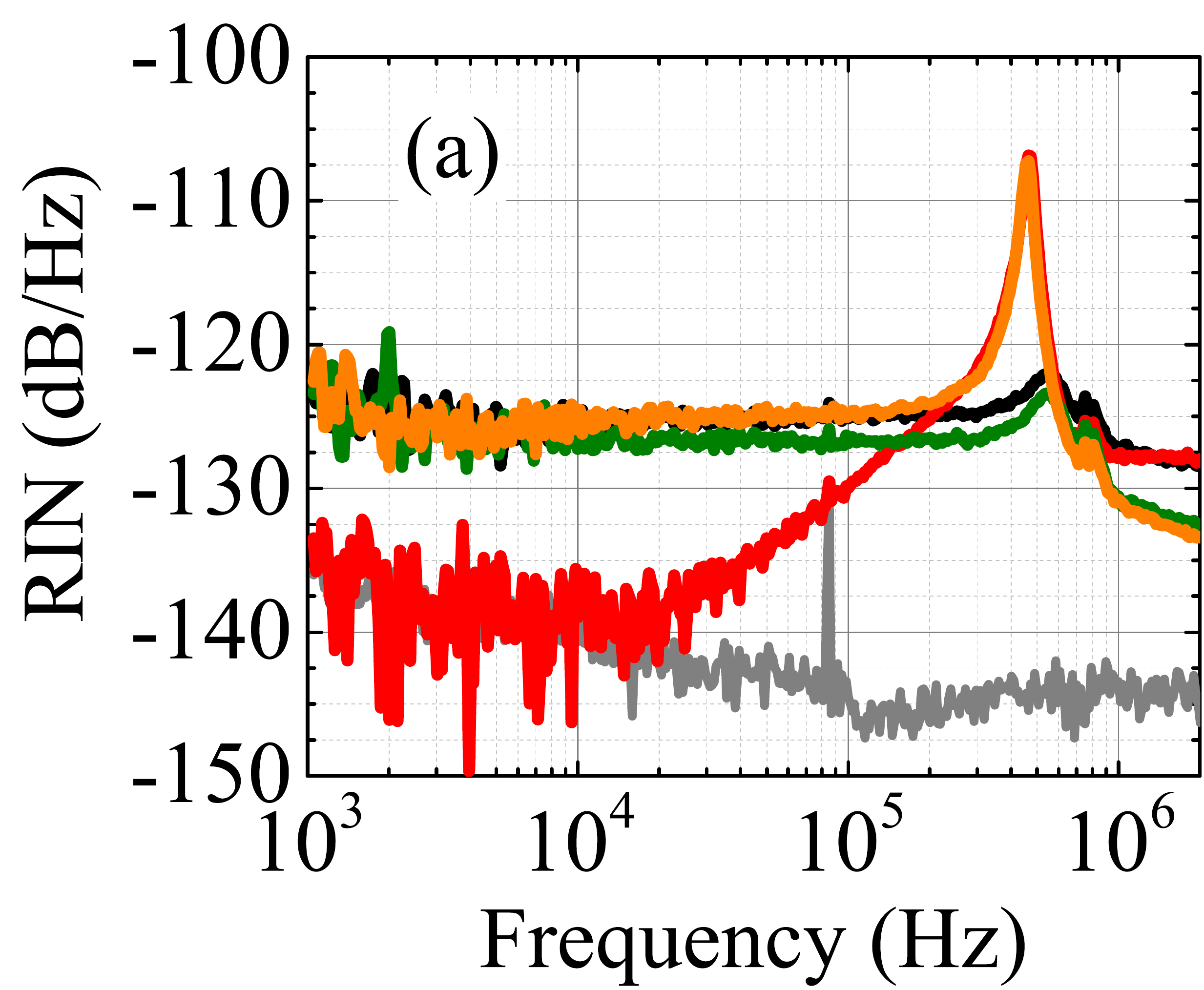} \includegraphics[width=.495\linewidth]{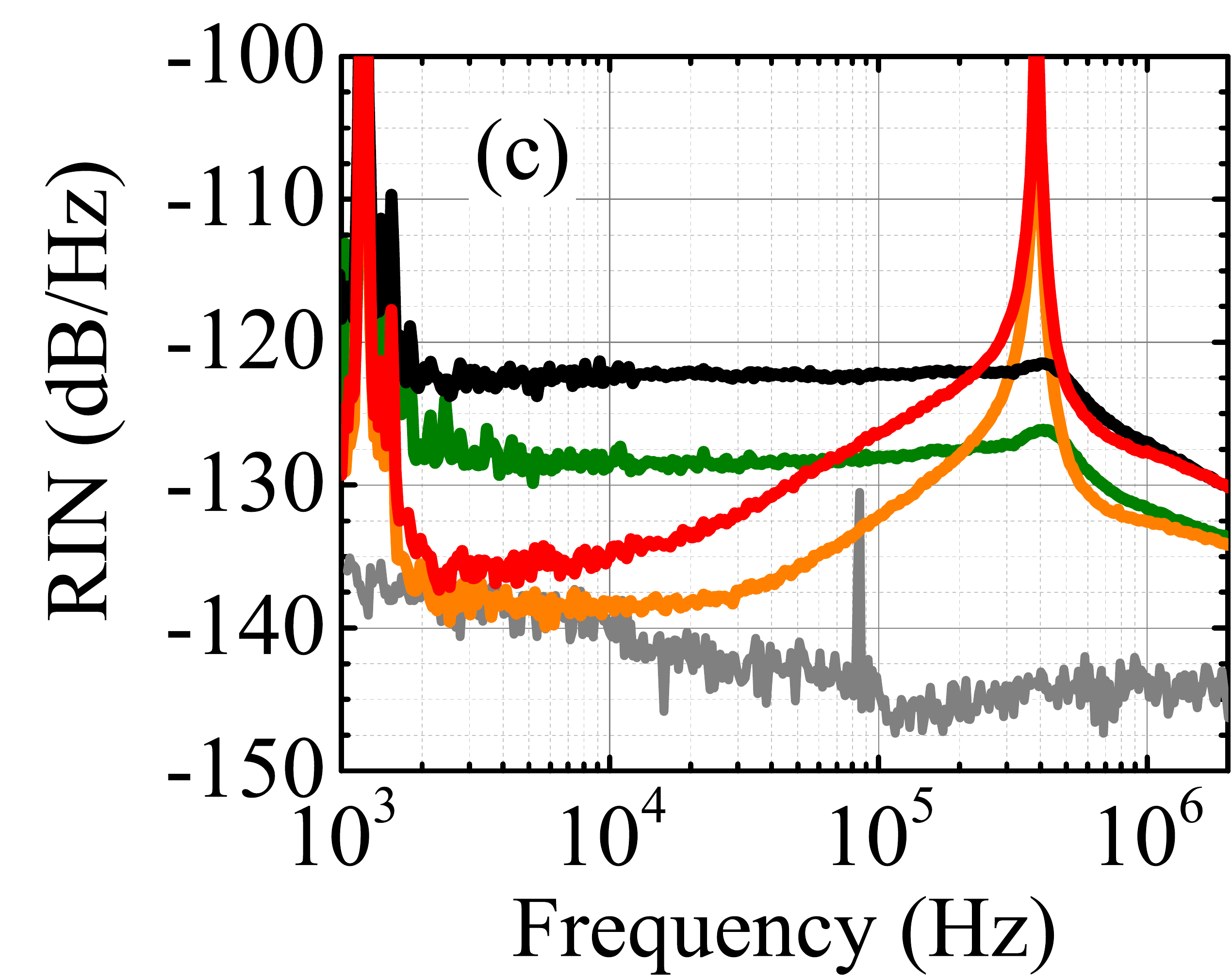}
 \includegraphics[width=.495\linewidth]{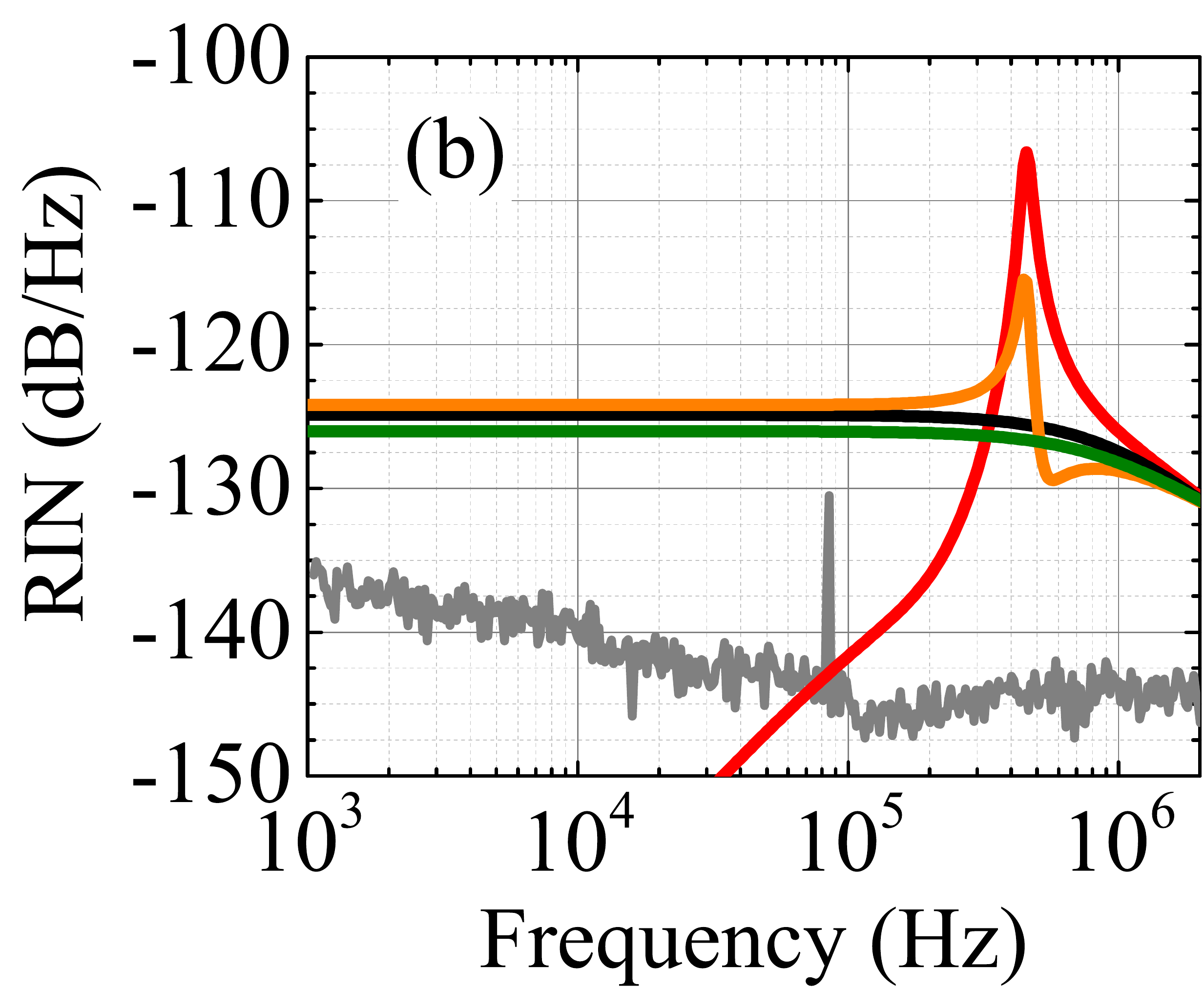}\includegraphics[width=.495\linewidth]{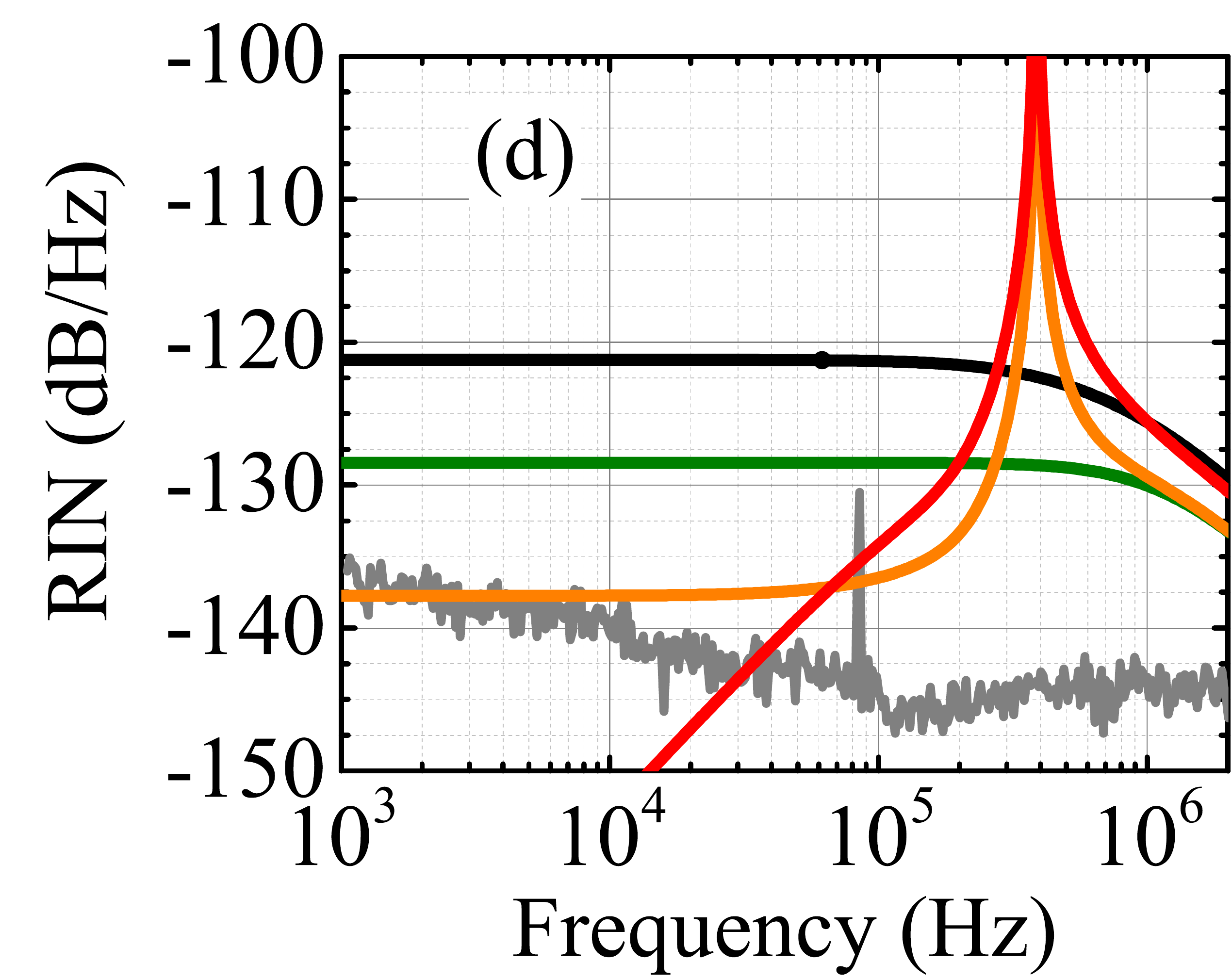}
\caption{Results of intensity stabilization based on detection of the $x$-polarized mode only. (a,c) Experimental and (b,d) theoretical RIN spectra. (a,b): pump configuration labeled \textcircled{\raisebox{-1pt}{1}} in Fig.\,\ref{fig1}. (c,d): pump configuration labeled \textcircled{\raisebox{-1pt}{2}}. Black (resp. green) line: $x$- (resp. $y$-) polarized mode free running RIN; Red (resp. orange) line: $x$- (resp. $y$-) polarized mode closed-loop RIN. Gray line: measurement floor. Loop parameters: $f_\mathrm{PI} = 200 \,\mathrm{kHz}$, $f_\mathrm{LP} = 250 \,\mathrm{kHz}$, $Q=0.8$ with $G=0.2$ for (b) and $G=0.075$ for (d). Laser parameters : $\mathrm{RIN_p}=-140\,\mathrm{dB/Hz}$, $r_x = 1.92$, $r_y=1.95$, $\tau_x= 18\,\mathrm{ns}$, $\tau_y= 14\,\mathrm{ns}$, $C=0.4$, $\eta=0.4$ for (b) and $\mathrm{RIN_p}=-133\,\mathrm{dB/Hz}$, $r_x = 2.1$, $r_y=1.55$, $\tau_x= 65\,\mathrm{ns}$, $\tau_y= 45\,\mathrm{ns}$, $C=0.05$, $\eta=0.98$ for (d). The shot--noise level is around -150\,dB/Hz.}
\label{fig7}
\end{figure}

An intensity servo--loop control has been experimentally implemented for the two pumping architectures labelled  \textcircled{\raisebox{-1pt}{1}} and \textcircled{\raisebox{-1pt}{2}} in Fig.\,\ref{fig1}. First, the feedback signal is extracted from detection of only one polarization, namely the $x$ polarization. 
Figures \ref{fig7}(a,c)  show the measured RINs of the $x$- and $y$-polarized modes for the two pumping architectures for free running and closed-loop operations. Figures \ref{fig7}(b,d) show the associated predictions based on the above model of Eqs.\,(\ref{eq:15}, \ref{eq:16}). A very good agreement is achieved between theory and experiment. The closed-loop signals display a peak after 400~kHz, which is due to the loop filter phase shift induced by the limited bandwidth of the  servo--loop. Figure \ref{fig7}(a) corresponds to the single pump beam configuration \textcircled{\raisebox{-1pt}{1}}, for which the laser parameters are $C=0.4$ and $\eta=0.4$. While the noise for the $x$-polarized mode is efficiently reduced by the servo-loop, a $3$~dB  noise degradation is observed on the cross-polarized mode. This is well reproduced by the model in Fig.\,\ref{fig7}(c) and was expected from the anti-phase correlations of the mode noises with such  parameters as shown in subsection \ref{sec:2}\ref{subsec:3A}. Indeed, the pump correlation amplitude $\eta=0.4$ is well below the value $\eta_\mathrm{inv}=0.9$ obtained from \eqref{eq:10} with the corresponding pumping parameters. To summarize, in this situation, stabilizing the intensity of one mode degrades the noise of the other one.


On the contrary, Fig.\,\ref{fig7}(c) is based on the dual pump beam configuration \textcircled{\raisebox{-1pt}{2}}, which leads to $C=0.05$ and $\eta=0.98$. The intensity noises of both modes are simultaneously suppressed by the servo-loop. This is well reproduced by the model in Fig.\,\ref{fig7}(d) and was expected from the fact that the intensity noises of the modes are in-phase correlated with these pumping parameters leading to $\eta_\mathrm{inv}=0.45$, as seen in subsection \ref{sec:2}\ref{subsec:3A}.

In both pumping architectures, we can notice a bump in the measured free-running spectra (black and green lines) at 500\,kHz, which is not reproduced by the model. This bump appears in fact in the intensity noise of the pump laser diode at high pump currents, whereas we assumed a white pump noise in our model.

Since the dual-pump beam architecture labeled \textcircled{\raisebox{-1pt}{2}} is found to be the best one for joint noise stabilization of the two modes and noise minimization at the same time, it will be the only one investigated in the following. 

\begin{figure}[h!]
\centering
\includegraphics[width=.4\textwidth]{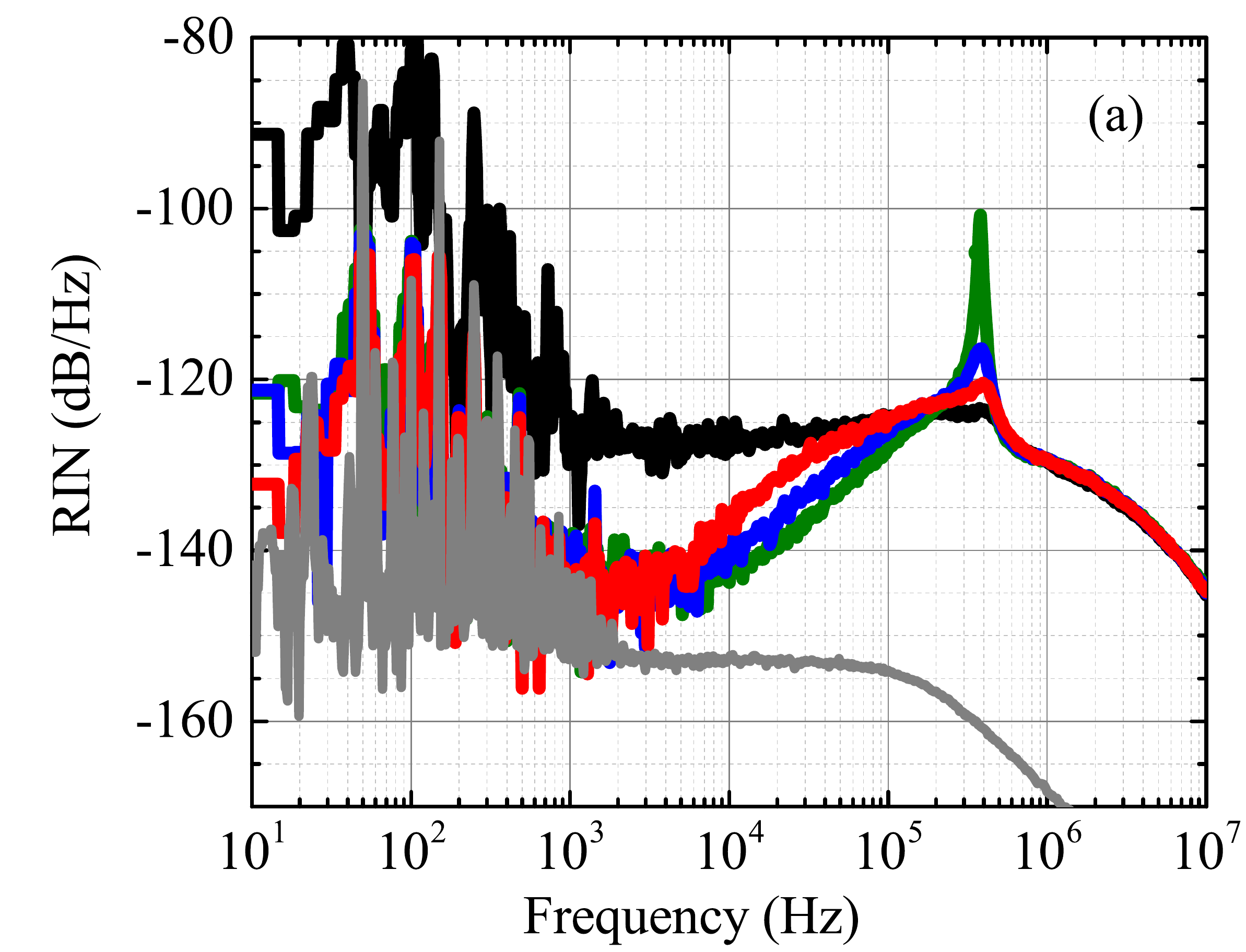} \includegraphics[width=.4\textwidth]{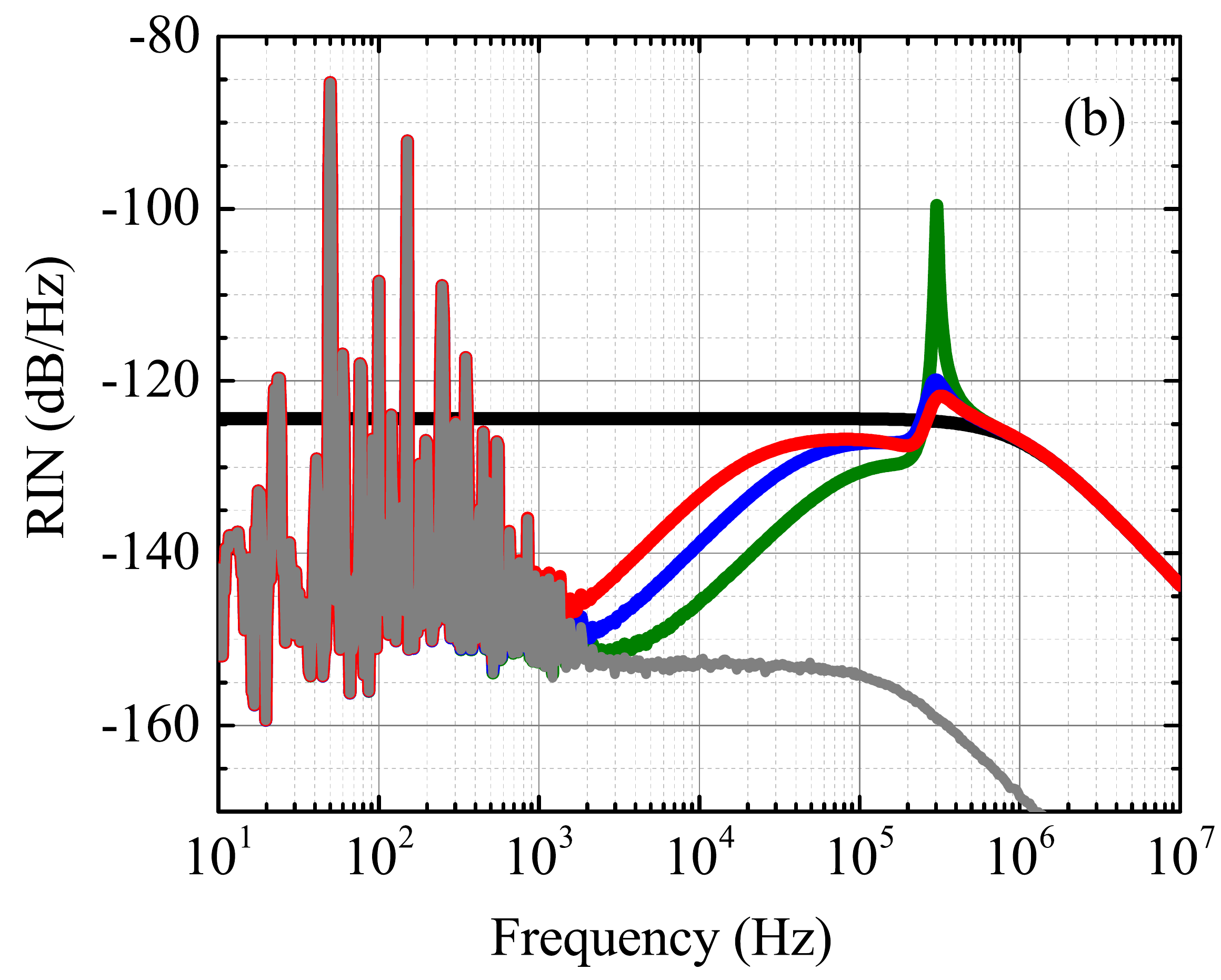}
\caption{Results of intensity stabilization based on detection of the total laser power using the in--phase fully correlated pumping architecture \textcircled{\raisebox{-1pt}{2}}. (a) Experimental and (b) theoretical RIN spectra. 
Black line: free-running laser RIN. Green line: RIN in closed-loop operation. Blue and red lines: Closed-loop RIN spectra for two different implementations of an extra  lead--lag filter. Gray line: measured electronic noise floor, used to model the RIN spectra of (b). Loop parameters: $f_\mathrm{LP}=250\,\mathrm{kHz}$, $f_\mathrm{PI}=200 \,\mathrm{kHz}$, $G=0.08$ and $Q=1$. Lead--lag filters parameters:  $\tau_\mathrm{peak} = 0.5 \,\mu\mathrm{s}$, $\tau_\zeta=0.9\,\mu\mathrm{s}$ (blue line), $\tau_\zeta =1.8\,\mu\mathrm{s} $ (red line). Laser parameters: $\tau=1\,\mathrm{ns}$, $r_x=1.38$, $r_y=1.23$, $C=0.05$, $\eta=0.98$,  $\mathrm{RIN}_\mathrm{p}=-133\,\mathrm{dB/Hz}$,  $\tau_x=30\,\mathrm{ns}$,   $\tau_y=17\,\mathrm{ns}$. The shot--noise level is around -153\,dB/Hz. }
\label{fig8}
\end{figure}

We now turn to the situation where the feedback signal is  extracted from the measurement of the total output power of the DF-VECSEL. The best intensity noise reduction of the total output signal is obtained in this configuration as reported in Fig.\,\ref{fig8}(a). The free running RIN for the total laser output power is reproduced as a black line and the locked one as a green line. A total RIN level below -140~dB/Hz is obtained below 10~kHz. The spurious peaks below 1~kHz come from technical noises, such as electronic noises, mechanical vibrations or thermal noise and should be cancelled using a proper isolation of the system. After locking, the intensity noise is dominated by the peak above 300~kHz, which is due to the limited bandwidth of the servo-loop and can be  eliminated by improving the filter, as is going to be investigated in the next section.   

\subsection{Improvement of the loop filter}\label{Leadlag}

The DC offset fluctuations of the commercial servo--controller that we use, which can be seen as gray line in Fig.\,\ref{fig8}, are responsible for the low-frequency (below 1~kHz) noise of the RIN of the intensity-stabilized DF-VECSEL. The power spectral density (PSD) $\mathcal{S}_\mathrm{DC}$ of this DC offset noise is thus introduced in the expression of the closed-loop RIN expression of \eqref{eq:15}, leading to:
 \begin{eqnarray}
\mathrm{RIN}^\mathrm{lock.}_{x+y}\left(f\right)  = &  \left|\dfrac{1}{1+H\left(f\right)\cdot \sum_{ij} M_{ij}\left(f\right)}\right|^2 \mathrm{RIN}_{x+y}\left(f\right)  \nonumber \\
& +  \left|\dfrac{H\left(f\right) \cdot \sum_{ij} M_{ij}\left(f\right) }{1+H\left(f\right)\cdot \sum_{ij} M_{ij}\left(f\right)}\right|^2  \cdot \dfrac{\mathcal{S}_\mathrm{DC}}{\mathrm{DC}^2} \,. \label{eq:18}
 \end{eqnarray}
Figure \,\ref{fig8}(b) shows that this model based on \eqref{eq:18} perfectly reproduces the measurements. This allows to  numerically optimized the parameters of the loop filter transfer function and to tailor the electronics to fit our specific needs. A lead-lag filter is thus implemented in order to suppress the noise peak above 300~kHz. This phase compensation filter is actually a simple passive circuit with a resistor ($R_1$)  in parallel with a resistor-capacitor ($R_2$, $C_2$) circuit \cite{nise:2011}. Its transfer function involves then two characteristic response times : $\tau_\mathrm{peak} = R_2\,C_2$ and $\tau_{\zeta} = C_2\,\left(R_1 + R_2\right)$. For the design of the filter, $\tau_\mathrm{peak}$ is adjusted to center the phase compensation at the peak frequency, while $\tau_{\zeta}$ allows to increase the damping ratio $\zeta$ of the quasi-oscillation. Such an analog lead-lag filter proves to be very efficient for the optimization of the locked RIN, as displayed in the red and blue lines of Fig.\,\ref{fig8}(a) for two different values of the parameters of the filter. Finally, using such a filter, the measured closed-loop RIN spectra lie below -120~dB/Hz and are well reduced at low frequencies.

 The question to know whether such intensity noise performances are good enough for application to the CPT-clock will be addressed in section \ref{sec:4}. Before that, we turn to the active stabilization of the RF beat-note phase noise, which is needed to reach a level compatible with good short-time stability performances for the clock. 

\section{Beat-note phase stabilization}  \label{sec:3}

Excitation of the ground state hyperfine transition of cesium relies on the RF beat-note between the $x$- and $y$-polarized modes. However, the beat-note generated by the DF-VECSEL exhibits a relatively large phase noise with a typical bandwidth of several hundreds of kilohertz. Previous studies have shown that, below a few tens of kilohertz, this phase noise is dominated by thermal fluctuations \cite{De:2015, Liu:2018, Gredat:2019}. For the sake of clarity, we adopt a simple macroscopic description of thermal fluctuations \cite{De:2015}, while further microscopic refinement have recently been developed \cite{Gredat:2019}. The relevant parameters that govern this noise mechanism in this simple approach are the pump powers   $P_{\mathrm{p},x}$ and $P_{\mathrm{p},y}$ of the two modes, the thermal resistance $R_\mathrm{T}$ and the thermal response time $\tau_\mathrm{T}$ of the 1/2--VCSEL, and its refractive index variation with temperature $\Gamma_\mathrm{T}$. Besides, the high--frequency component of the beat--note phase noise comes from the conversion of the intensity noise of the two modes into optical phase noise due to the large Henry factor $\alpha$ of such lasers. The pumping architecture  \textcircled{\raisebox{-1pt}{2}} has proven to be efficient for minimizing the pump-to-laser intensity noise transfer, thus limiting the bandwidth of the free running beat-note phase noise to about 100~kHz \cite{Gredat:2018}. Such a limited noise bandwidth allows us to consider the implementation of an optical phase-locked loop (OPLL) to mitigate the beat-note phase noise, as is now going to be investigated. 

\begin{figure}[h]
    \centering
    \includegraphics[width=.99\linewidth]{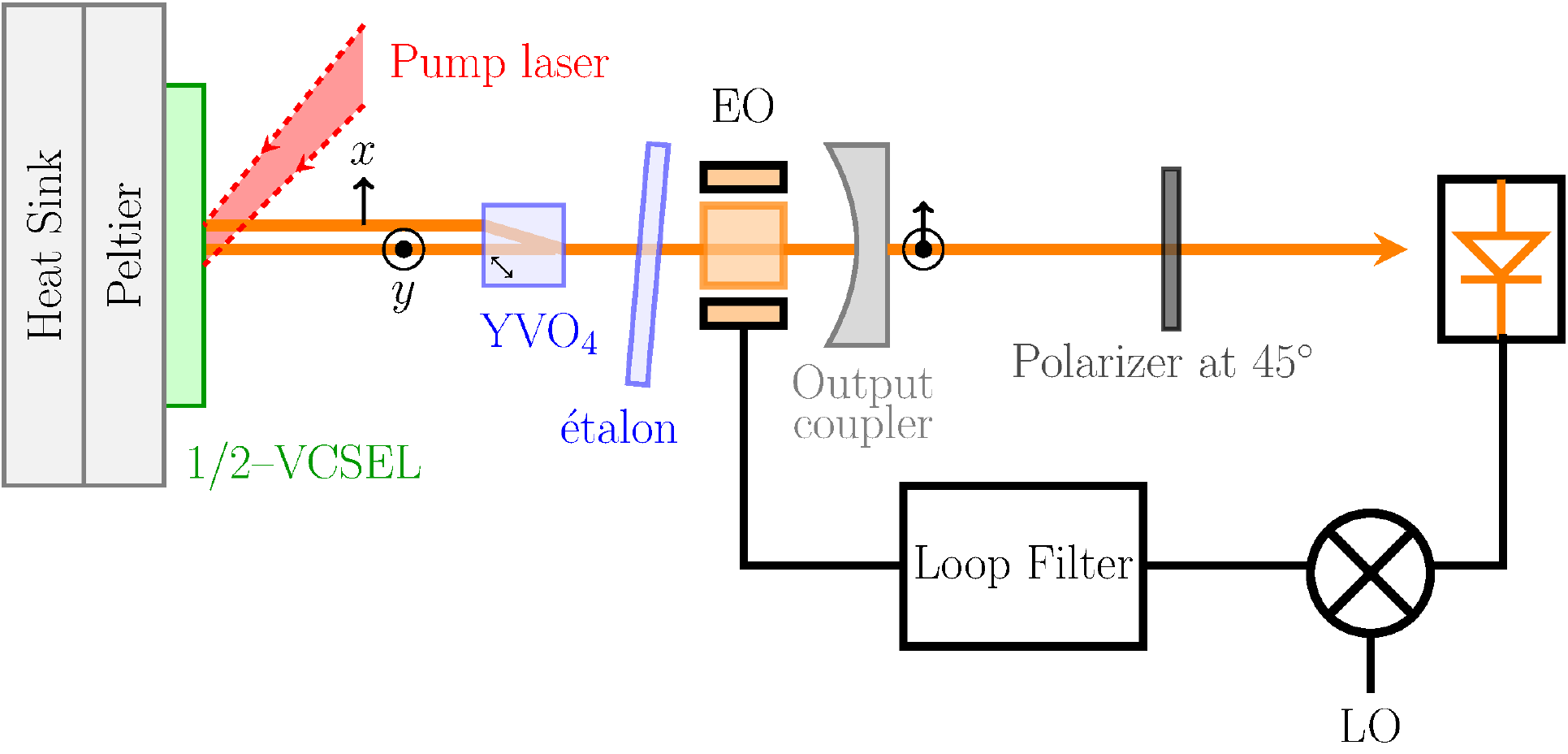}  
    \caption{Schematics of the OPLL controlling the beat-note phase noise. EO: intra-cavity electro--optic MgO:SLT crystal. LO: radiofrequency local oscillator to which the DF-VECSEL beat-note phase is locked. }
    \label{fig9}
\end{figure}

\subsection{Optical phase-locked loop}

The active stabilization of the beat-note phase requires to control the intra-cavity birefringence using an electro--optic (EO) crystal, as sketched in Fig.\,\ref{fig9}. Stoichiometric lithium tantalate (SLT) is used for the EO crystal because of its large electro-optic coefficients $r_{33}$ and $r_{13}$ along the axes $x$ and $z$, where $x$ is the axis along which the voltage is applied and $z$ denotes the beam propagation direction. This crystal is doped with MgO in order to reduce photorefractive effects. The dimensions of the MgO:SLT used in the present study are $2 \times 2 \times 1 \,\mathrm{mm}^3$. The DF-VECSEL beat-note signal is detected after the beam passes through a polarizer oriented at $45^\circ$ of the $x$ and $y$ directions. After mixing with an RF local oscillator to downshift the frequency, the phase noise is filtered to create an error signal, which is then applied as a feedback voltage to the EO crystal.

\begin{figure}[h]
    \centering
    \includegraphics[width=.4\textwidth]{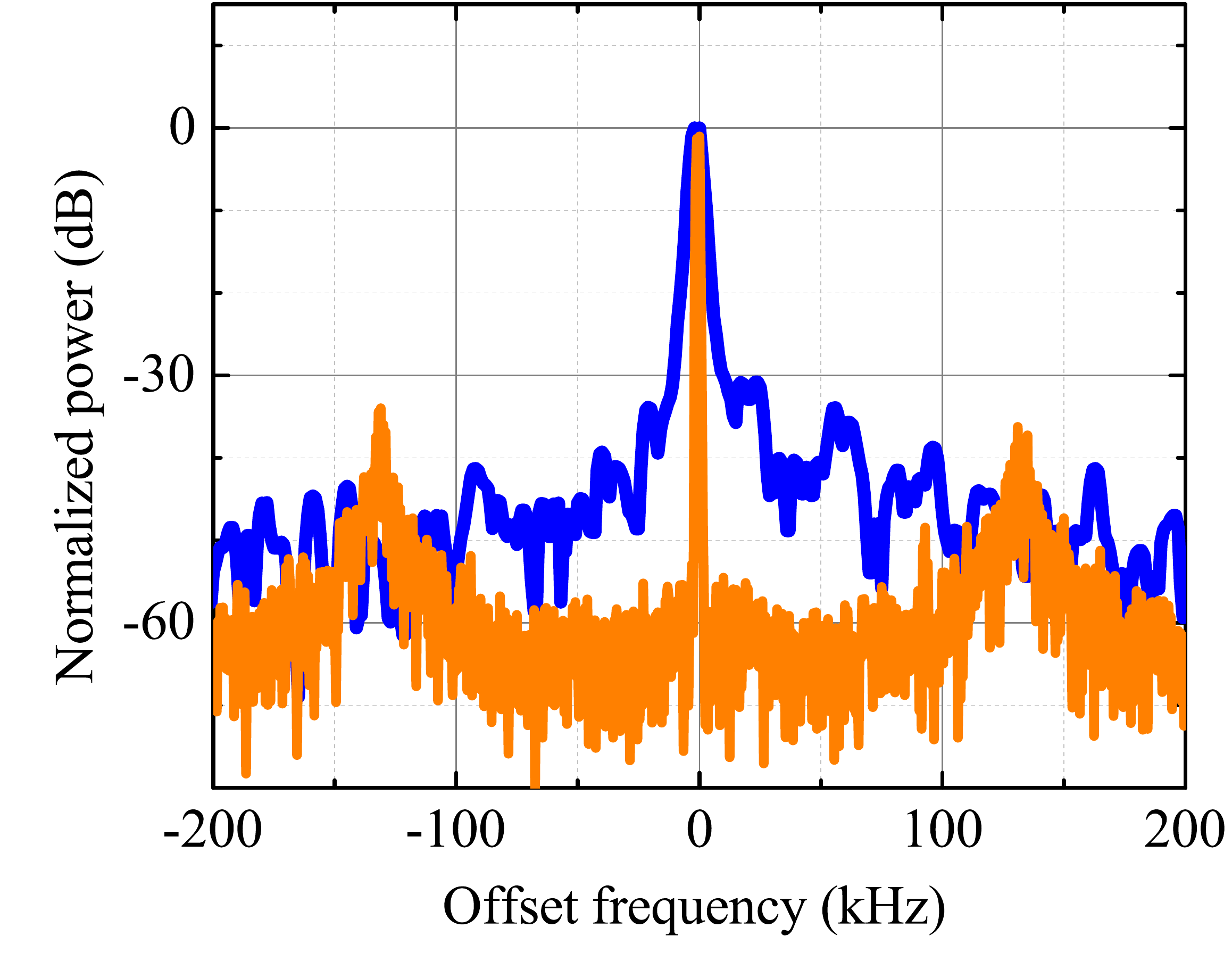}  
    \caption{Measured laser beat-note spectra in free running operation (blue line) and with the closed OPLL (orange line). Respective resolution bandwidths are 7.2~kHz and 820~Hz.}
    \label{fig10}
\end{figure}

Figure \ref{fig10} shows the effect of this OPLL on the beat-note spectrum. The free-running spectrum (blue line) exhibits a large phase noise pedestal, which is strongly reduced when the OPLL is closed (orange line). The OPLL bandwidth is larger than 100~kHz. The details of the OPLL leading to these results are given in section \ref{subsec:3C} below. They are based on the model detailed in the following section.

\subsection{OPLL Model}

\begin{figure}[h]
    \centering
    \includegraphics[width=.45\textwidth]{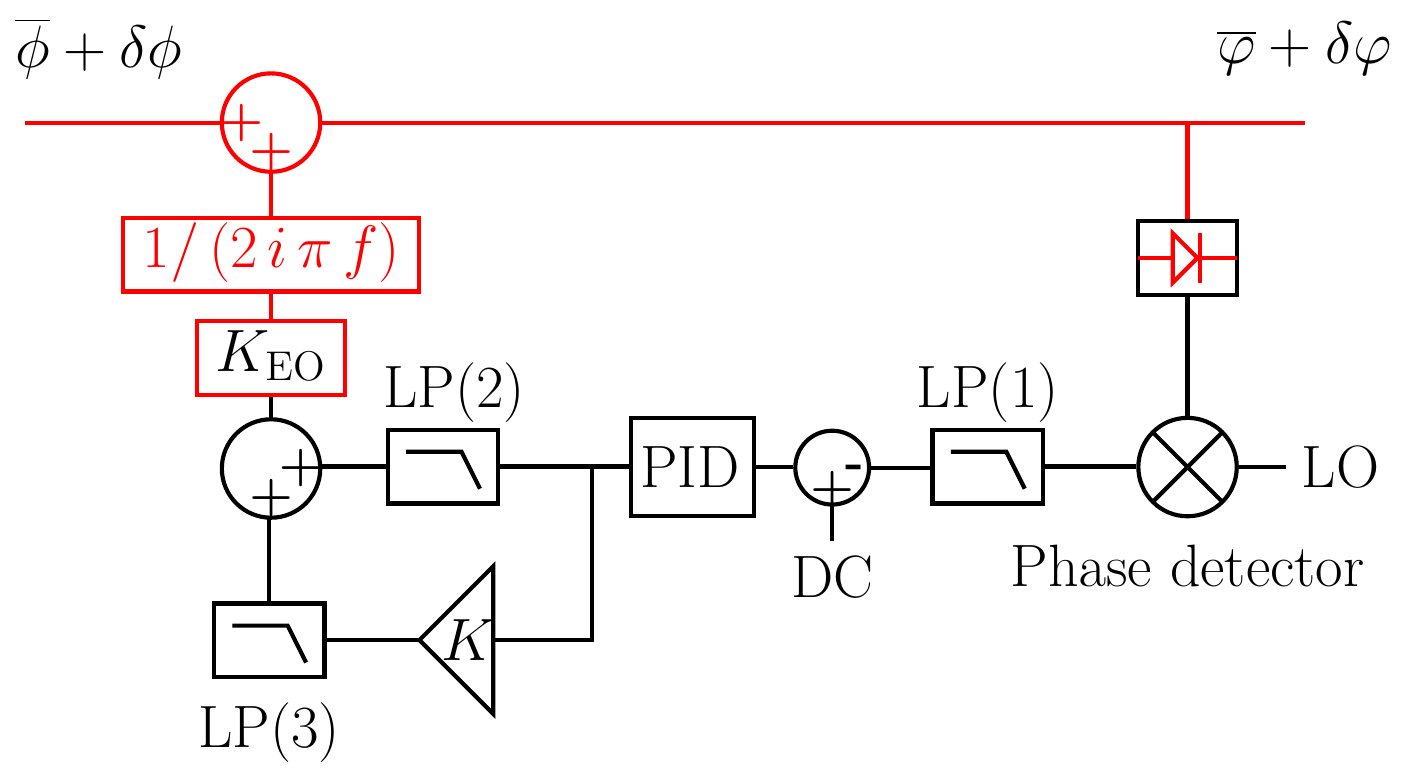}  
    \caption{Block diagram of the OPLL. $\phi$ (resp. $\varphi$):  free-running (resp. closed loop) beat-note phase. LP(1), LP(2), LP(3): low-pass filters. LO: local oscillator. }
    \label{fig11}
\end{figure}
Figure \ref{fig11} shows a block diagram of the OPLL. After photodetection, the beat-note signal is mixed with a local oscillator whose phase noise PSD is denoted as $S_{\phi_\mathrm{LO}}\left(f\right)$. After amplification and low-pass filtering (LP(1) in Fig.\,\ref{fig11}), the error signal is filtered by the servo-controller. The loop filter transfer function for unitary gain is denoted as $H_\mathrm{PID}$ and is characterized by the corner frequencies $f_\mathrm{PI}$, $f_\mathrm{D}$ and the derivative gain $G_\mathrm{D}$. As sketched in Fig.\,\ref{fig11}, two paths are used for the feedback on the EO crystal. One is direct (fast loop) while the other one goes through a high voltage amplifier (slow loop). The EO crystal has a capacitance $\varepsilon \,l_y\,l_z/l_x \simeq 0.4 \,\mathrm{pF}$, with the dielectric permittivity $\varepsilon$, and the crystal dimensions $l_x$, $l_y$, and $l_z$. The fast and slow loops can be typically considered as resistor-capacitor circuits and thus modeled as low--pass filters denoted as LP(2) and LP(3), respectively, in Fig\,\ref{fig11}. The total servo-loop transfer function $H$ for the phase noise can be expressed as~:
 \begin{equation}
 \label{eq:20}
H\left(f\right)= G \cdot H_\mathrm{PID}\left(f\right) \left[  H_\mathrm{LP(2)}\left(f\right)+ K \cdot H_\mathrm{LP(3)}\left(f\right)\right]  H_\mathrm{EO}\left(f\right)\,,
\end{equation}
where $K$ is the high voltage amplifier gain and $G$ a global gain factor. In \eqref{eq:20}, $H_\mathrm{EO}$ is the EO crystal voltage--to--phase transfer--function. When an electric field is imposed along $l_x$, this crystal changes the intra-cavity phase retardance between the ordinary and extra-ordinary polarizations characterized by their refractive indices $n_\mathrm{o}$ and $n_\mathrm{e}$, respectively. Inside the resonant cavity of round-trip  length $2\,L_\mathrm{cav}$, this retardance modifies the frequency difference between the $x$- and $y$-polarized modes. This leads to the voltage to frequency transfer function denoted as $K_\mathrm{EO}$ in Fig.\,\ref{fig11}. Once modified into a voltage--to--beat-note--phase transfer function, $K_\mathrm{EO}$ becomes \cite{Dumont:2014} : 
\begin{equation}
H_\mathrm{EO}\left(f\right) = \dfrac{1}{2\,\mathrm{i}\,\pi\,f}\cdot \dfrac{c}{2\,L_\mathrm{cav}}\cdot \dfrac{2\,\pi\,l_z}{\lambda}\cdot \dfrac{1}{l_x}  \left[\dfrac{n_\mathrm{e}^3 \cdot r_{33}}{2}-\dfrac{n_\mathrm{o}^3 \cdot r_{13}}{2} \right]\,.
\end{equation}
The resulting closed-loop beat--note phase noise PSD is related to the free running one $\mathcal{S}_\phi\left(f\right)$ through the following expression~: 
\begin{equation}
    \mathcal{S}^\mathrm{lock}_\varphi \left(f\right)=  \dfrac{\mathcal{S}_\phi\left(f\right) + \left|H\left(f\right)\right|^2 \cdot S_{\phi_\mathrm{LO}}\left(f\right)}{\left|1 + H\left(f\right) \right|^2} \,.
\end{equation}

\subsection{Experimental OPLL performances} \label{subsec:3C}

\begin{figure}[h!]
\centering
\includegraphics[width=.4\textwidth]{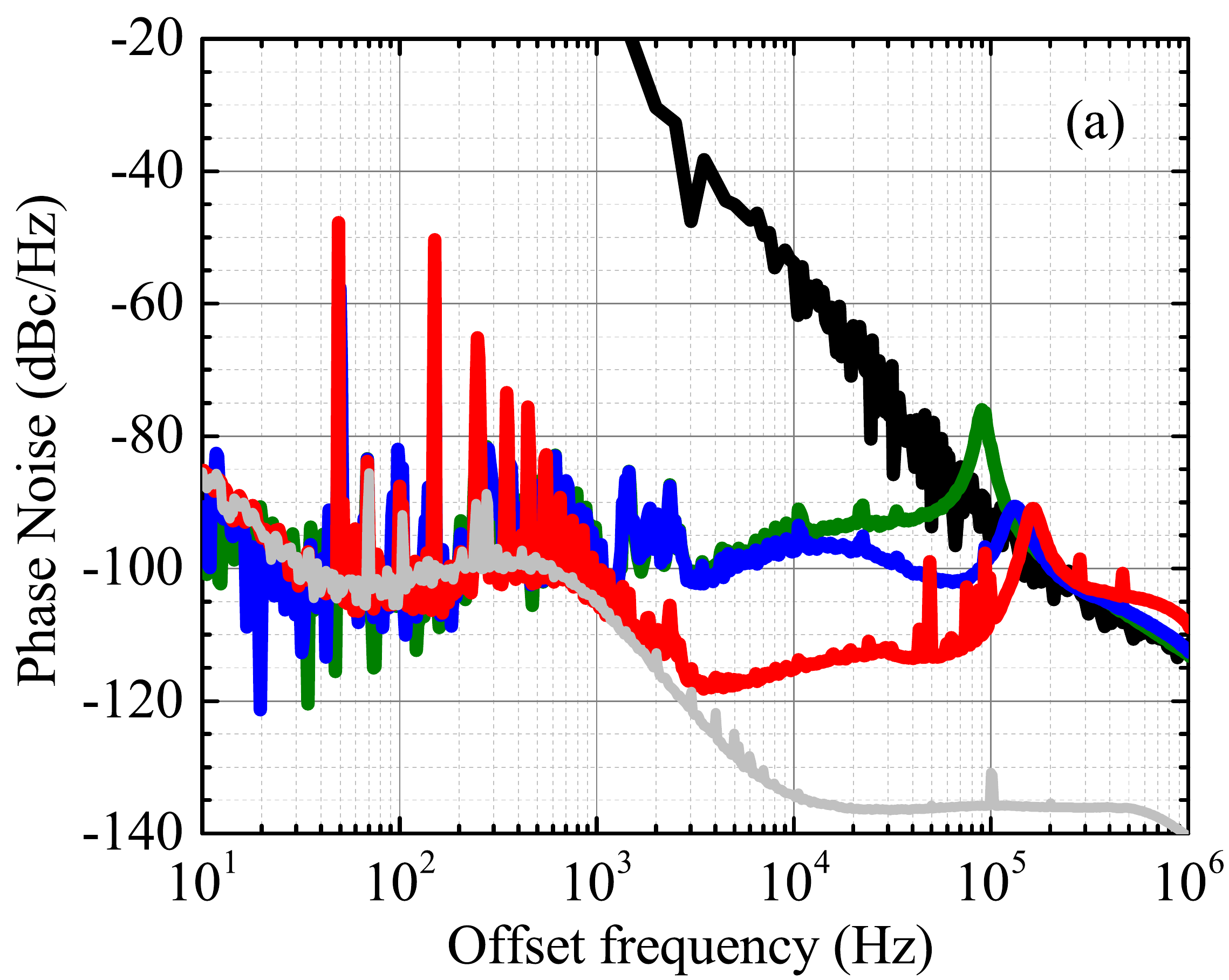}
\includegraphics[width=.4\textwidth]{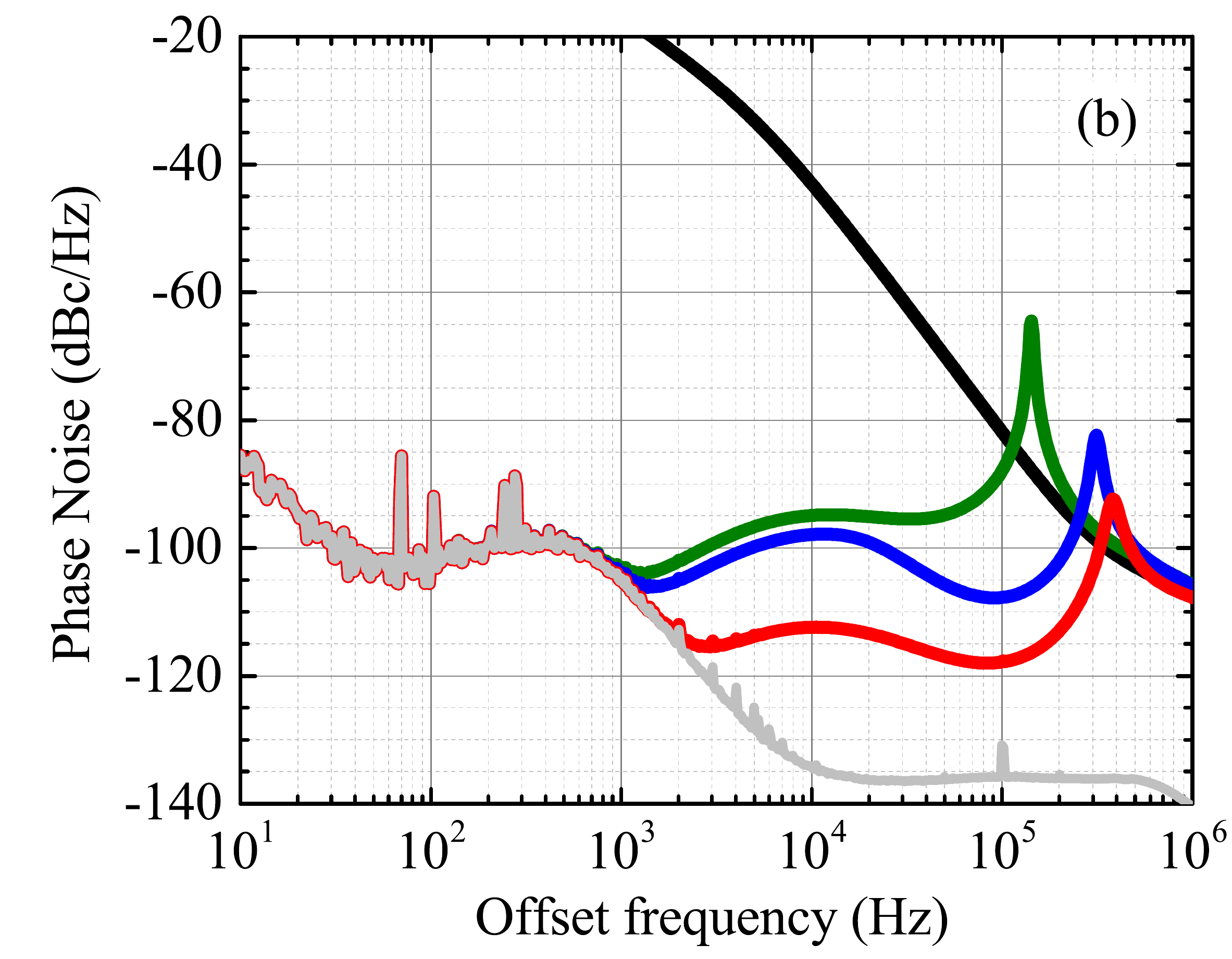}
\caption{(a) Measured and (b) computed free-running and closed loop beat-note single-sideband phase noise spectra. Black line: free-running laser. Green, blue and red lines: closed-loop phase noise spectra obtained with, respectively, a PI correction, a PID correction and a PID with maximum gain. Gray line: LO phase noise. Loops parameters: $f_\mathrm{PI}=20\,\mathrm{kHz}$, $f_\mathrm{D}=50\,\mathrm{kHz}$, $G_\mathrm{D}=15$, $K=30$. Laser parameters: $L_\mathrm{cav}=5\,\mathrm{cm}$, $\alpha=5.2$, $P_{\mathrm{p},x}=0.48\,\mathrm{W}$, $P_{\mathrm{p},y}=0.45\,\mathrm{W}$,  $R_\mathrm{T}=40\,\mathrm{K.W^{-1}}$, $\tau_\mathrm{T}=30\,\mu\mathrm{s}$,  $\Gamma_\mathrm{T}=1.39\times10^{-7}\,\mathrm{K^{-1}}$. The  other parameters have the same values as in Fig.\,\ref{fig8}. EO crystal parameters: $n_\mathrm{e} = 2.1418$, $n_\mathrm{o}=2.1441$, $r_{33} = 6.96\,\mathrm{pm.V^{-1}}$, $r_{13} = 29.6 \,\mathrm{pm.V^{-1}}$ \cite{Casson:2004, Schossig:2014,Dumont:2014}.}
\label{fig12}
\end{figure}

The black line in Fig.\,\ref{fig12}(a) shows the free running beat-note phase noise spectrum obtained with the pumping architecture labeled \textcircled{\raisebox{-1pt}{2}} in Fig.\,\ref{fig1}. Above 100\,kHz, the phase noise PSD is lower than -95~dBc/Hz thanks to the noise minimization allowed by this pumping configuration \cite{Gredat:2018}. The other spectra in Fig. \ref{fig12}(a) show that a tremendous beat-note phase noise reduction is achieved within this 100\,kHz bandwidth when we implement the OPLL. A level as low as -103 dBc/Hz at 100 Hz frequency is demonstrated after locking. This value is clearly limited in our case by the noise of the local oscillator (gray line). First, the green line corresponds to the case where the correction used is a simple proportional-integral. The experimental noise spectrum is well reproduced by the model, as shown by the green line in Fig. \,\ref{fig12}(b). To get a better noise suppression, a differential correction is added so as to further reduce the phase noise in the intermediate frequency range between 20~kHz and 100~kHz, as shown by the blue lines in Fig.\,\ref{fig12}(a) and (b). Finally, by further increasing the OPLL gain, still better noise performances are reached for the locked beat-note phase noise, as one can notice from the red lines in Figs.\,\ref{fig12}(a) and (b).
 
Now that both intensity and beat-note phase noise performances in the presence of their respective optimized stabilization loops have been investigated, we can, in the following section, evaluate the achievable CPT clock short-time stability with such an actively stabilized DF-VECSEL. 

\section{Prediction of clock short-time stability}  \label{sec:4}

The relative frequency stability of the CPT clock is deteriorated by the probe laser fluctuations \cite{Danet:2014} during the atomic interrogation process, owing to the so-called Dick effects \cite{dick:87}. To predict the clock performances, three contributions of the DF-VECSEL fluctuations are considered \cite{Dumont:2014} : i) the RIN, ii) the RF beat-note phase noise, and iii) the fluctuations of the laser optical frequency around the atomic transition frequency at 852~nm. A relative Allan deviation of the clock below $\sigma_y=5\times10^{-13}$ at one second averaging time is targeted in the present work. The reduction of the third contribution has already been addressed thanks to the active stabilization of the optical frequency of a 852~nm VECSEL \cite{Dumont:2014}, so that we do not consider it here. We therefore focus on the two other contributions for which our active stabilization efforts are presented in sections \ref{sec:3} and \ref{sec:4}. Given the atoms interrogation sequence developed at Observatoire de Paris-SYRTE, which is detailed in \cite{Danet:2014}, and the atomic response, the contributions of the intensity noise and beat-note phase noise can be readily evaluated  \cite{Dumont:2014}. The Ramsey-like clock sequence has a total duration $T_\mathrm{c} = 6\,\mathrm{ms}$ and pulse shaping is obtained thanks to acousto-optic modulators. This sequence is composed of a first
pumping pulse of duration $\tau_\mathrm{p} = 2\,\mathrm{ms}$, a free-evolution time $T_\mathrm{R} = 4\,\mathrm{ms}$, and a very short detection pulse lasting $\tau_\mathrm{d} = 25\,\mu\mathrm{s}$. The power transmitted by the cesium cell is measured during this latter detection step. 
\begin{table}[h]
\centering
\caption{Clock short--time stability contribution of the DF-VECSEL intensity noise and beat-note phase noise. Allan deviations are numerically computed in the pumping configuration \textcircled{\raisebox{-1pt}{2}} and the data come from Fig.\,\ref{fig8}(a) and Fig.\,\ref{fig12}(a). In the part using the model, fixed values for all the parameters are kept while $\eta$ is varied. The part computed from the measurements involves the interpolation of the measured noise spectra at the harmonics of the interrogation frequency $1/T_\mathrm{c} \simeq 167\,\mathrm{Hz}$ over a 10\,MHz bandwidth.} 
\begin{tabular}{|l|c|c|}
\hline
 & Intensity noise & Phase noise  \\ 
\hline
Situation &  \multicolumn{2}{c|}{Dick effect $\sigma_y  \times 10^{13}$ at 1 s}\\
\hline
\multicolumn{3}{|c|}{From the model without stabilization} \\
\hline   
     Free running with $\eta =0$                       & $22$  & $8.3 \times 10^4$ \\
     Free running with $\eta =0.5$                      & $20$ & $5.9 \times 10^4$\\
     Free running with $\eta =1$                      & $17$ & $3 200$ \\
 \hline
\multicolumn{3}{|c|}{From measurements} \\
 \hline
 Free running & 69 & $\times$ \\
 RIN locked & 5.8 & $\times$ \\
 RIN locked with filter (1) &  4.1 & $\times$ \\
 RIN locked with filter (2) & 4.2 & $\times$ \\
 \hline
 OPLL PI & $\times$ & 14\\
 OPLL PID & $\times$ & 9.1\\
 OPLL PID (Max Gain) & $\times$  & 2.3 \\
LO & $\times$ & 1.7 \\
 \hline
\end{tabular}
\label{tab}
\end{table}

Taking the Dick effect into account, the relative Allan deviations $\sigma_y$ due to intensity noise or beat-note phase noise are numerically evaluated either with the noise spectra predicted from our model or directly from the measurements. The results are gathered in Table\,\ref{tab}, first for the free-running laser followed by the stabilized laser. For the free-running laser, the fully in--phase correlated pumping \textcircled{\raisebox{-1pt}{2}} of Fig.\,\ref{fig1}, leading to a strong reduction of the laser noise, is shown to  decrease the contribution of the beat-noise phase noise to the Allan deviation by a factor of 26. 

Table\,\ref{tab} also shows that the predicted Allan deviation due to the free-running intensity noise can be more than 3 times lower when computed from the model rather than  computed from the measurements. The aforementioned spurious peaks in the intensity noise are responsible for this discrepancy since they are not reproduced by the model. As a consequence, using only the noise modelling to evaluate the Allan deviation can give a trend but the prediction of the clock short time stability should rely on the measurements.

Considering now the best experimental measurements we obtained for the phase and intensity stabilized laser, i.~e., the blue line of Fig.\,\ref{fig8}(a) for the RIN  with filter (1) and the red line of Fig.\,\ref{fig12}(a) for the phase noise, the results of Table\,\ref{tab} lead to the following numerical evaluation of the total Allan deviation for the clock : 
\begin{equation}
    \label{eq:23}
    \sigma_y^\mathrm{Dick}\left(1\,\mathrm{s}\right) = \sqrt{\left(4.1\right)^2+\left(2.3\right)^2}\times 10^{-13} < 5 \times 10^{-13}\,.
\end{equation}
An Allan deviation of $4.7\times 10^{-13}$ is thus expected, which is compatible with the targeted performances. Furthermore, a great part of the beat-note phase noise contribution comes in fact from the local oscillator phase noise, as one can notice from the last line of Table\,\ref{tab}. The use of a dedicated optimized local oscillator could further improve this performance \cite{Francois:2014}.

\section{Conclusion}

In conclusion, we have shown that DF--VECSELs are good candidates as laser sources to improve the trade-off performance vs size of cesium CPT clocks provided a careful active stabilization of their intensity and beat-note phase is implemented. In particular, a thorough understanding of the influence of mode competition on the correlations of the noises of the two laser modes has been shown to be mandatory to optimize the active stabilization loops. In particular, in--phase correlation of the intensity fluctuations of the two modes is required to achieve mutual reduction of the RINs of the two modes. The associated pumping architecture, using two separated pump beams, is also very important to reach a low noise beat--note. Thanks to the implementation of two feedback loops and their optimization, a short--time relative frequency stability below the targeted value of $\sigma_y\left(1\,\mathrm{s}\right)~=~5\times~10^{-13}$ can be reached. A model has been established for the laser in the presence of its active electronic servo-loops, which exhibits a very good agreement with the experiments and allows to point out possible limits and improvements of the stabilization loops. Finally, an interesting extension of the present work would be to get rid of the external local oscillator by using an optical delay line to reduce the beat--note phase noise \cite{Pillet:08}, thus turning the DF-VECSEL into a more compact self-stabilized optoelectronic oscillator.

\section*{Funding Information}

Direction G\'en\'erale de l'Armement; Agence Nationale de la Recherche (CHOCOLA, ANR-15-CE24-0010-04, DIFOOL ANR-15-ASMA-0007-04, LASAGNE, ANR-16-ASTR-0010-03); joint lab between LAC and Thales R\&T.



\bigskip

\bibliography{sample}




\end{document}